\documentclass[twocolumn]{aastex62}
\usepackage{amsmath,amstext}
\usepackage{savesym}
\savesymbol{tablenum}
\usepackage{apjfonts} 
\usepackage{verbatim}
\usepackage{chngcntr}
\usepackage{siunitx}
\usepackage{placeins}
\newcommand{\be}{\begin{eqnarray}}
\newcommand{\ee}{\end{eqnarray}}

\begin{document}

\normalsize


\title{\Large \textbf{Tidal Heating of Exomoons in Resonance and Implications for Detection}}

\author{Armen Tokadjian}
\affiliation{Department of Physics and Astronomy, University of Southern California, Los Angeles, CA 90089-1342, USA; tokadjia@usc.edu}
\affiliation{The Observatories of the Carnegie Institution for Science, 813 Santa Barbara St., Pasadena, CA 91101, USA}

\author{Anthony L. Piro}
\affiliation{The Observatories of the Carnegie Institution for Science, 813 Santa Barbara St., Pasadena, CA 91101, USA}

\begin{abstract}
The habitability of exoplanets can be strongly influenced by the presence of an exomoon, and in some cases the exomoon itself could be a possible place for life to develop. For moons outside of the habitable zone, significant tidal heating may raise their surface temperature enough to be considered habitable. Tidal heating of a moon depends on numerous factors such as eccentricity, semimajor axis, size of parent planet, and presence of additional moons. In this work, we explore the degree of tidal heating possible for multi-moon systems in resonance using a combination of semi-analytic and numerical models. This demonstrates that even for a moon with zero initial eccentricity, when it moves into resonance with an outer moon, it can generate significant eccentricity and associated tidal heating. Depending on the mass ratio of the two moons, this resonance can either be short-lived ($\leq200$ Myr) or continue to be driven by the tidal migration of the moons. This tidal heating can also assist in making the exomoons easier to discover, and we explore two scenarios: secondary eclipses and outgassing of volcanic species. We then consider hypothetical moons orbiting known planetary systems to identify which will be best suited for finding exomoons with these methods. We conclude with a discussion of current and future instrumentation and missions.

\end{abstract}


\section{Introduction}

Exomoons play an important role in planetary systems. By impacting the tidal heating and geology of the bodies in the system, they can affect habitability and astrobiology. A moon can stabilize a planet's tilt \citep{laskar1993}, prevent tidal locking between host and star \citep{Tokadjian}, drive tidal heating (\citealp{dobosheller,Piro18}), sustain a planet's magnetic field \citep{andrault}, and has the potential to host life itself. This could occur when a moon sits within the circumstellar habitable zone of the parent star or when the moon is sufficiently heated to generate an alternative habitable zone (\citealp{reynolds,scharf,hellerarmstrong}).

Despite their importance, exomoons remain challenging to detect. Some of the methods discussed for detecting exomoons include transit timing variations (\citealp{hek,heller}), direct imaging with spectroastrometry \citep{agol}, and microlensing \citep{liebig}. Unfortunately, the small size of rocky exomoons makes applying these methods difficult with current instrumentation. Nevertheless, the example of Io in our own solar system hints at additional strategies for potentially discovering the first rocky exomoons. Due to the compact arrangement of the inner moons of Jupiter, Io is strongly tidally heated (\citealp{peale1979,peters,oza,rovira}). This makes it a bright infrared source (e.g. \citealp{depater}, \citealp{skrutskie}, and references therein), and causes venting of volcanic species \citep{lellouch1990,lellouch1996}. Given the large number of satellites of the gas and ice giants in our solar system, it is natural to expect similar planets across the galaxy will be home to multi-moon systems \citep{sasaki10}, some of which will also result in strong heating from tidal interactions.

Motivated by these issues, in this work we study the tidal interactions of multi-moon systems both semi-analytically and numerically to explore the range of tidal heating that may be possible in extrasolar planetary systems. In analogy to dynamics experienced by Io, we focus on bodies in mean motion resonance, MMR, and apply these ideas to the restricted 3-body problem of two moons orbiting a planet.

 By comparing theory to simulation, we explore how a moon in orbital resonance undergoes libration in its semimajor axis and eccentricity, which affects the amount of tidal heating it receives (\citealp{pealecassen,peale1979,Henning}). We compare these simulations to three body systems like Jupiter, Io, and Europa, and demonstrate that even with a parameterized model for heating we are able to replicate the eccentricity and temperature of Io. We then extend this same model to known exoplanets and consider the presence of hypothetical moons orbiting in mean motion resonance. This further illuminates the dependence of heating on factors such as planet composition and orbital period (\citealp{peters,oza}). As a moon's temperature increases, it opens new possible routes for detection through techniques such as secondary transits and volcanic outgassing. We explore the possibility of observing exomoons using these methods. 
 
This study is organized as follows. In Section~\ref{sec:resonance}, we lay out the framework for the restricted three body problem in mean motion resonance and compare numerical simulation to theoretical work. Next, we introduce tidal heating in Section~\ref{sec:heating} and apply our model to a variety of exomoon configurations, as well as compare with the Jupiter-Io-Europa system. This is extended to exoplanet systems with hypothetical moons in Section~\ref{sec:exoplanets} where we also explore the observational aspects for possibly detecting these moons. Finally, we conclude with a summary in Section~\ref{sec:conclusion}.

\section{Orbital Resonance}
\label{sec:resonance}

We begin by summarizing some of the main features of the dynamics of a system consisting of a parent planet orbited by two moons near resonance. This will set the stage for our later calculation of the tidal heating due to these dynamics. We do this by focusing on the restricted three body problem where a central mass is orbited by two entities: a massless test particle and a perturbing body (\citealp{bruno,goldreichs}). For our analysis, we further simplify the problem by assuming the central planet and the outer moon do not undergo orbital evolution, but the test particle moon is subject to changes in semimajor axis and eccentricity. Generally speaking, when the two moons are near a mean motion resonance, specifically with orbital periods in the ratio of 2:1, the changes in these orbital parameters depend on the degree of resonance, or how close the system is to perfect resonance.

In this section, we first include a basic introduction to the theoretical background of the restricted three body problem and mean motion resonance. Next we present numerical simulations of these phenomena, discuss the agreement with theory, and highlight the additional details that numerical simulations provide.

\subsection{Resonance in the Restricted Three Body Problem}
\label{3body}
Mean motion resonance of celestial bodies in orbit around a primary has been studied in depth, with numerous applications ranging from solar system objects \citep{peale} to specific exoplanetary systems \citep{correia}. Here we mostly follow the mathematical formulism set forth by \citet{petrovich}.

\begin{figure}
\epsscale{1.0}
\plotone{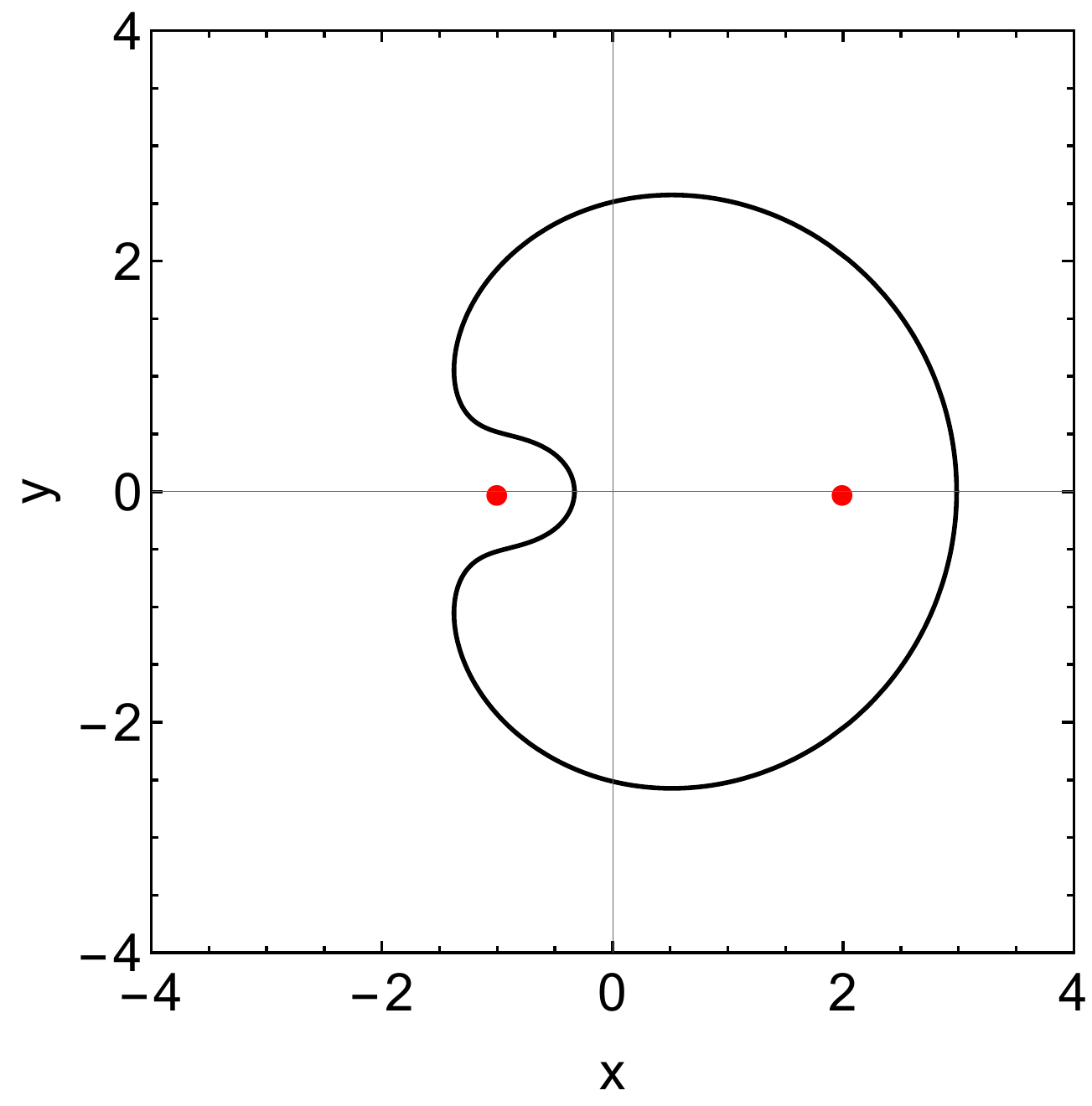}
\caption{A constant Hamiltonian curve for a planet orbited by a test particle moon and perturbing moon in a 2:1 resonance. Here $\Delta = 1$ and $K=0.5$. The $x$ and $y$ coordinates represent the canonical momentum and canonical coordinate, functions of semimajor axis and eccentricity that vary for all points except the fixed points. These are shown here in red at (-1,0) and (2,0).}
\label{Kcontheory}
\epsscale{1.0}
\end{figure}

Our system consists of a planet as the primary, a test particle moon in orbit around the planet, and an outer perturbing moon in a 2:1 resonance with the inner moon. Given this resonance, we have a relation between the period of the moons,
\be
    \frac{P_2}{P_1}=\frac{2}{1},
    \label{12res}
\ee
where $P_1$ and $P_2$ are the orbital periods of the inner and outer moon, respectively. The Hamiltonian of the system is derived for this restricted three body problem and can be written,
\be
    K = -3\Delta R + R^2 - 2\sqrt{2R}\;\rm{cos}(r)
    \label{Hamiltonian}
\ee
(see \citealp{petrovich} for full details of the calculation). Here $\Delta$ is the resonance distance,
\be
    \Delta = 0.4\left(\frac{m_p}{m_2}\right)^{2/3}\left[2\left(\frac{a_1}{a_2}\right)^{5/6}-\frac{6+5e_1^2}{6+9e_1^2}\left(\frac{a_2}{a_1}\right)^{2/3}\right],
    \label{resdis}
\ee
for planet mass $m_p$, inner moon with semimajor axis $a_1$ and small eccentricity $e_1$, and perturbing moon of mass $m_2$ and semimajor axis $a_2$. $R$ is the canonical momentum,
\be
    R = 0.9\left(\frac{m_p}{m_2}\right)^{2/3}\left(\frac{a_2}{a_1}\right)^{2/3}\left(\frac{6e_1^2}{6+7e_1^2}\right),
    \label{canmom}
\ee
and $r$ is the canonical coordinate,
\be
    r = -2\lambda_2+\lambda_1+\omega_1,
    \label{cancoord}
\ee
where $\lambda_1$ and $\lambda_2$ are the mean longitude of the inner and outer moon, respectively, and $\omega_1$ is the longitude of periapsis of the inner moon. For constant Hamiltonian $K$, the semimajor axis and eccentricity of the inner moon will oscillate, or librate, about certain equilibrium values. These values are obtained by minimizing the Hamiltonian which translates to solving the cubic equation,

\be
    x^3-3\Delta x - 2 = 0,
    \label{minK}
\ee
where $x$ is the canonical momentum in Cartesian coordinates,
\be
    x = \sqrt{2R} \;\mathrm{cos}(r).
    \label{xcoord}
\ee
Together with $y$, the canonical coordinate in Cartesian coordinates,
\be
    y = \sqrt{2R} \; \mathrm{sin}(r),
    \label{ycoord}
\ee
these coordinates track the orbital oscillations for constant $K$, as shown in Figure~\ref{Kcontheory}. In this case, arbitrary values of $0.5$ and $1$ are chosen for $K$ and $\Delta$, respectively, with $\Delta$ representing the shape of the moon trajectory in the constant Hamiltonian phase space. Because these values do not correspond to a minimum of $K$, there is libration of the orbital parameters, as can be seen by the bean-shaped trajectory. The libration period can be estimated to be \citep{murray}

\be
    t_{\mathrm{lib}} = 0.4P_1\left(\frac{a_2}{a_1}\right)^{2/3}\left(\frac{m_p}{m_2}\right)^{2/3}\left(1+\frac{5}{6}e_1^2\right),
    \label{tlib}
\ee
where $P_1$ is the orbital period of the inner moon. This shows that as the planet mass dominates the perturbing moon or we move to larger MMR ratios, there is slower change in the orbital values of the system.

Thus, the closer to true resonance, the smaller the Hamiltonian of the system, and as it approaches a minimum, the semimajor axis and eccentricity vary less and less \citep{murray}. Conversely, the Hamiltonian can be increased to obtain a greater variation in orbit, while the moons are still considered to be in resonance. Nevertheless, for any degree of resonance we find a non-zero eccentricity is inevitable, which motivates our later calculations of tidal heating. In the example above, if the semimajor axis and eccentricity were chosen such that $K$ was minimized, we would end up in one of the fixed points (-1,0) or (2,0) shown in red, and there would be no oscillation in these parameters (no libration). In other words, the system would be in perfect resonance.

\begin{figure}
\epsscale{1.0}
\plotone{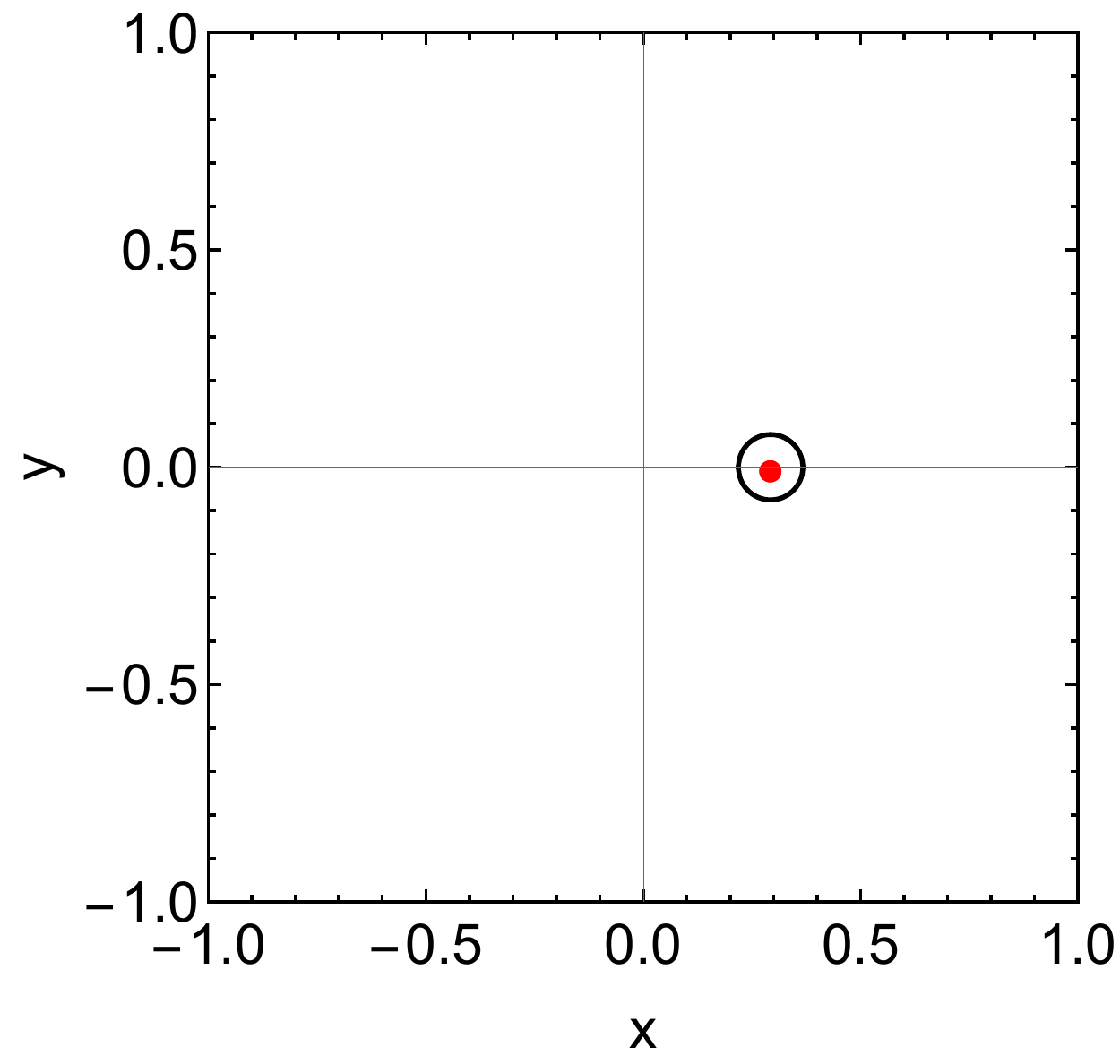}
\caption{A constant Hamiltonian curve for the Jupiter-Io-Europa system. Here $\Delta = -2.245$ and $K=-0.276$. Because this system is close to perfect resonance, there is little libration and the trajectory is a small circle. The minimum of the Hamiltonian, the fixed point, is the red point at (0.293,0), which is in the center of the trajectory.}
\label{JupIoTheo}
\epsscale{1.0}
\end{figure}

A well-known example of a system close to perfect resonance is Jupiter orbited by Io and Europa with orbital periods of 1.77 days and 3.55 days, respectively. Although the mass of Io is not negligible compared to the mass of Europa (a ratio of about 1/2) so that the restricted three body problem is not directly applicable, we show that the formulas described so far can still be used to approximate the trajectory (Figure~\ref{JupIoTheo}). For this system, $\Delta=-2.245$ and $K=-0.276$ and because we are close to the Hamiltonian minimum, the phase curve is simply a small circle around the fixed point. Thus, although we are not strictly in the restricted three body regime, the method can still be applied for a test particle that is small compared to central body.

The equations above apply to two moons in the specific 2:1 resonance configuration, but they can be easily extended to apply to other first order resonances with minor changes (see \citealp{petrovich} for details). Thus, because 2:1 mean motion resonance is simplest and most common to exoplanet systems \citep{lissauer}, we choose to focus on this configuration. We will briefly discuss the effect other orders of resonance have on tidal heating in the Section~\ref{cplheating}.

\subsection{Simulations of Resonance using REBOUND}
\begin{figure}
\epsscale{1.15}
\includegraphics[width=8cm,height=5cm]{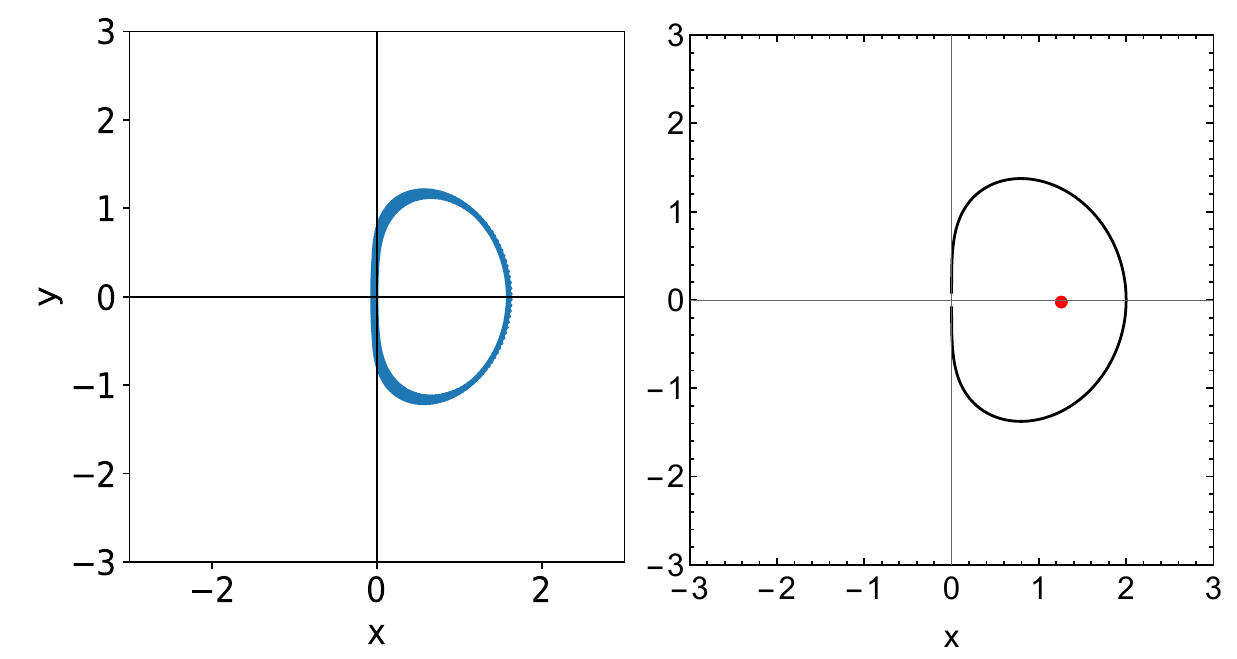}
\caption{Constant Hamiltonian phase curves of a planet orbited by two moons in resonance and zero initial eccentricity for simulation (left) and theory (right). Here $m_p/m_2 = 300$ and the inner and outer moons have orbital periods of 2 days and 4 days, respectively. The trajectory shows that the eccentricity does not remain zero but oscillates along with semimajor axis. The fixed point is the red dot at (1.26,0) and corresponds to the minimum of the Hamiltonian.}
\label{zeroecc}
\epsscale{1.0}
\end{figure}

Due to the complicated nature of even a three body system under gravitational influence, we utilize the code REBOUND \citep{rebound}, which is an N-body integrator that solves the motion of particles under the force of gravity. We first consider the zero eccentricity case, and take a generic three body system consisting of a planet and two moons. The mass ratio of planet to outer moon is 300, and the moons are placed in a 2:1 resonance. Integrating forward in time using REBOUND, we obtain the trajectory shown in the left panel of  Figure~\ref{zeroecc}. Here $\Delta=0$ and $K=0$ because of the initial zero eccentricity, the location of this initial configuration is the origin of the $xy$ plot. The right panel of the figure shows the theoretical calculation of this system (as described in Section~\ref{3body}), which is similar, but not identical, since the simulated phase curve is smaller. This is due to the fact that REBOUND takes into account the center of mass of the system that the bodies orbit. The analytic model, on the other hand, does not include these effects but has the center of mass located exactly at the origin of the coordinate system, the center of the planet. To test this explanation, we ran additional simulations for this system but with increasing mass ratio $m_p/m_2$. The result was a trend toward better agreement between theory and simulation for higher mass ratio, confirming that the discrepancy is due the location of the center of mass. Nonetheless, the theory and numerics agree well even for low mass ratio, demonstrating the robustness of the model.

\begin{figure}
\epsscale{1.1}
\plotone{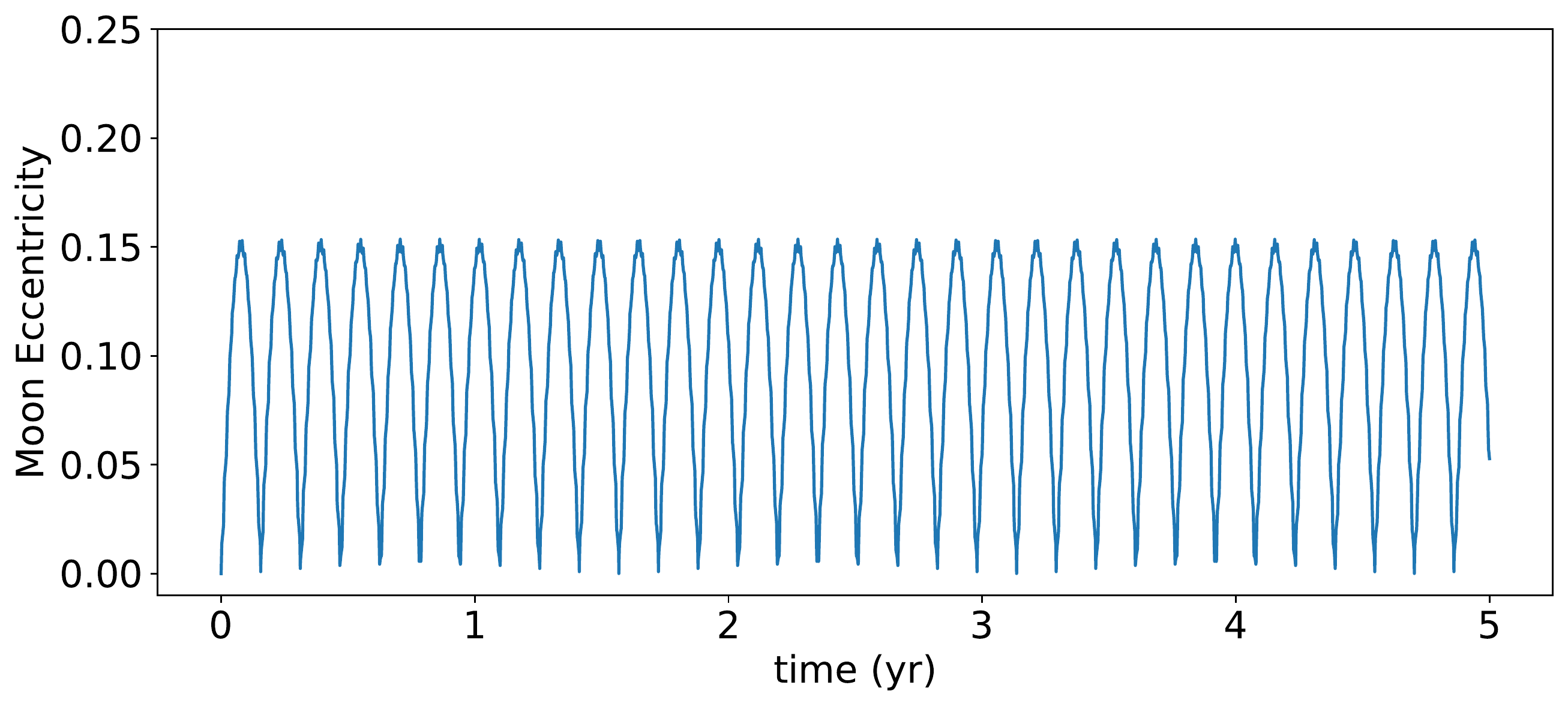}
\caption{Inner moon eccentricity evolution over time for the system in Figure~\ref{zeroecc}. The initially circular orbit grows in eccentricity and then oscillates with an amplitude of 0.1536 due to the system's distance from the minimum of the Hamiltonian.}
\label{eccosc}
\epsscale{1.0}
\end{figure}

The trajectories in Figure~\ref{zeroecc} show that the eccentricity, although initially set to zero, does not remain zero but oscillates about an equilibrium value (Figure~\ref{eccosc}). We can also adjust the Hamiltonian to obtain greater variation as we move away from the fixed point. Thus, by adjusting the energy of the system through parameters like semimajor axis, we can control the amplitude of libration. This has implications for tidal heating as discussed in the next section. 

\begin{figure}
\epsscale{1.0}
\plotone{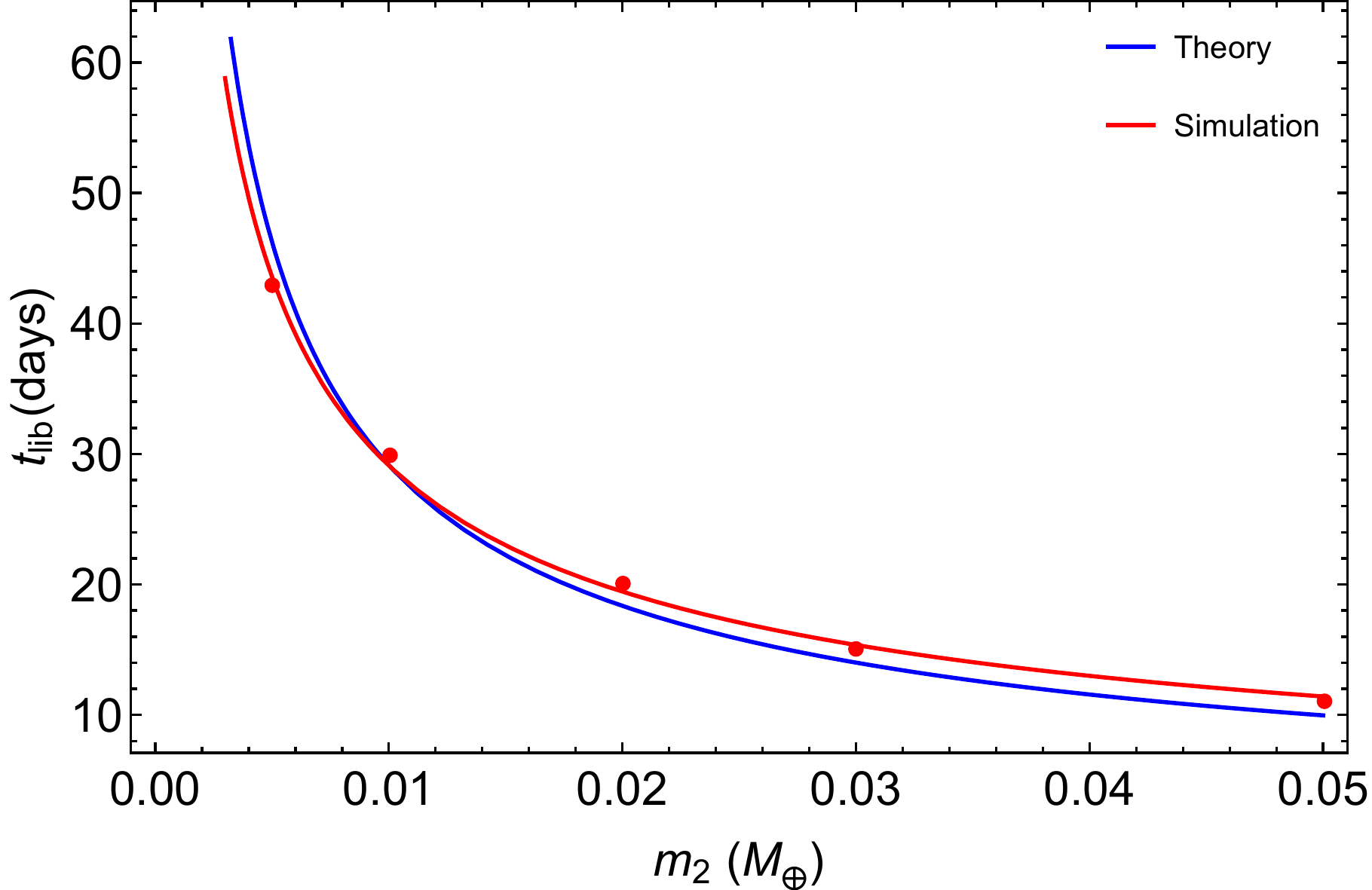}
\caption{Libration period versus perturbing mass for the theoretical relation given by Equation~(\ref{tlib}), in blue, compared to simulation results, shown as red points. The mass of the planet is $1.37 M_{\oplus}$ and the eccentricity of the inner moon orbit is 0.094. The red curve is the best fitting power law to the simulation results with a power law index of 0.73.}
\label{tlibcomp}
\epsscale{1.0}
\end{figure}

In addition to amplitude, the period of libration can be chosen based on the relation given in Equation~(\ref{tlib}). With a set planet mass and set distances such that a 2:1 resonance is achieved, the oscillation period scales with $m_2^{-2/3}$. We can check the validity of this power law by comparing the libration times we get for various outer moon sizes through simulation. We choose a planet mass of $1.37M_{\oplus}$ and an inner moon eccentricity of $0.094$, and plot the time given by Equation~(\ref{tlib}) alongside the simulation results for five different perturbing masses (Figure~\ref{tlibcomp}). The agreement is not exact, but the power law of each differ by less than $10\%$ ($0.67$ for theory, $0.73$ for simulation). The difference can be attributed to the changing eccentricity in the simulation due to being near-resonance, whereas the eccentricity was fixed for the theoretical result.

\section{Tidal Heating}
\label{sec:heating}

The previous section demonstrated that librations in eccentricity is an inevitable result of an inner moon being perturbed by an outer moon close to resonance. The time varying force converts some of the orbital energy into heating the moon (e.g., \citealp{kaula,peale1979,ogilvietide}), and we explore the strength of this tidal heating next. In the following subsections, we first describe the tidal heating model we employ and apply it to the basic case of Jupiter and its two inner moons. We then explore the parameter space of planet composition and orbital period, and their impact on heating.

\subsection{The CPL Model of Heating}
\label{cplheating}

The constant phase lag (CPL) model  describes the response of a body to a gravitational tug based on the angle, or phase, between the tidal bulge and the line joining the centers of the bodies (e.g. \citealp{efroimsky13}). There is a delay between where the planet exerts a gravitational force and where the tidal bulge on the moon appears, and the phase difference is inversely proportional to the quality factor $Q$ that is related to how strongly the body dissipates orbital energy via tides. Although in general the tidal response of a moon can be more complicated than the CPL model, we use this parameterization because it allows us to explore a wide range of system parameters efficiently. For this parameterization, the tidal luminosity is given by

\be
    L_{\mathrm{tide}} = \frac{21k_2 G^{3/2}m_p^{5/2}R_1^5e_1^2}{2Q_1 a_1^{15/2}},
    \label{tidalL}
\ee
where $k_2$ is the Love number, $R_1$ is the radius, and $Q_1$ is the quality factor of the inner moon, and where we neglect terms of higher order in eccentricity (see \citealp{jackson}, \citealp{Henning}, and \citealp{peters}). Here, we also assume that the moon is tidally locked to its parent planet \citep{murray}. Taking the tidal flux to be the luminosity divided by the surface area of the moon and then using the Stefan-Boltzmann law, we obtain for the instantaneous effective temperature of the moon,

\be
    T_{\mathrm{tide}} = \left(\frac{L}{4\pi \sigma_{\mathrm{SB}}R_1^2}\right)^{1/4}.
    \label{tidalT}
\ee
where $\sigma_{\mathrm{SB}}$ is the Stefan-Boltzmann constant.

\begin{figure}
\epsscale{1.1}
\plotone{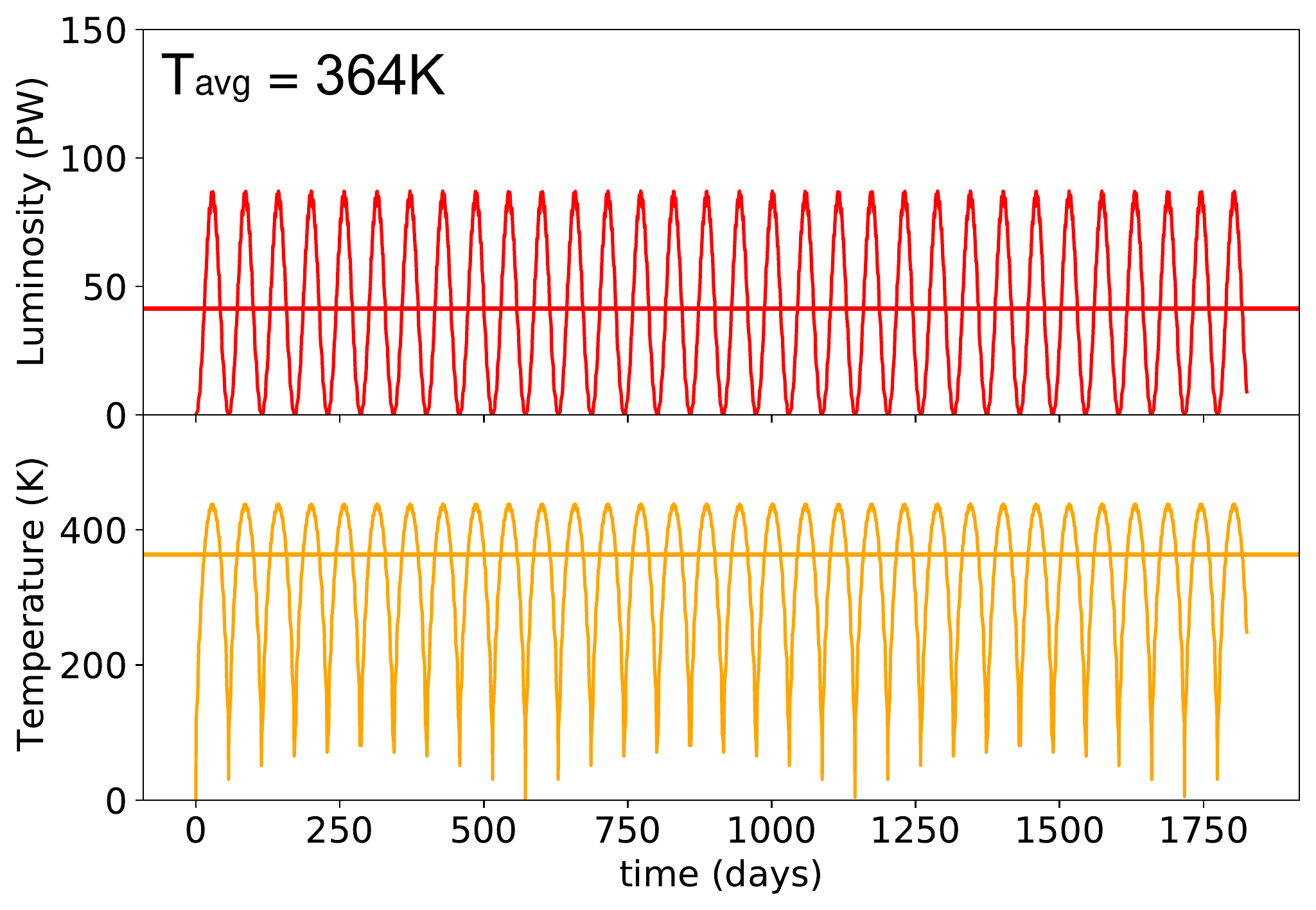}
\caption{Tidal heating luminosity and temperature for the system described in Figure~\ref{zeroecc}, where $m_p=4.5M_{\oplus}$. The horizontal red line shows the average luminosity from which the average temperature, the horizontal orange line, is calculated with $T_{\mathrm{avg}}$ shown at the top left.}
\label{zeroecctides}
\epsscale{1.0}
\end{figure}

Returning to the case shown in Figure~\ref{eccosc}, we simulate the same system in REBOUND and calculate the luminosity and temperature from tides using Equation (\ref{tidalL}) and (\ref{tidalT}) and present the results in Figure~\ref{zeroecctides}. Because the timescale for the heat transport across the moon is much longer than the period of librations, the surface temperature of the moon will reflect the average rather than the instantaneous heat flux. Taking 
\be
L_{\mathrm{avg}} = \frac{\int_0^{t_{\mathrm{lib}}}L_{\mathrm{tide}}\,dt}{t_{\mathrm{lib}}},
\ee
we then solve for 
\be
    T_{\mathrm{avg}} = \left(\frac{L_{\mathrm{avg}}}{4\pi \sigma_{\mathrm{SB}}R_1^2}\right)^{1/4}.
    \label{tidalTavg}
\ee
$L_{\mathrm{avg}}$ and $T_{\mathrm{avg}}$ are shown by the horizontal red and orange lines, respectively, with $T_{\mathrm{avg}}$ displayed in the top left corner.

The eccentricity at an exact resonance can be analytically estimated for a fixed mass ratio and semimajor axis ratio using Equations (\ref{minK}) and (\ref{xcoord}) for a constant Hamiltonian $K=0$. Solving Equation~(\ref{minK}) for $x$ with $K$ and $\Delta$ both zero, we get $x=2^{1/3}$. Then, plugging this into Equation~(\ref{xcoord}) with $r=0$, a canonical momentum of $R=2^{-1/3}$ is obtained. Inserting this into Equation~(\ref{canmom}) and solving for the eccentricity, we derive 
\be
    e_{1,\mathrm{res}} = 0.8\left[\left(\frac{m_p}{m_2}\right)^{2/3}-0.76\right]^{-1/2},
    \label{eres}
\ee
and for $m_p\gg m_2$,
\be
    e_{1,\mathrm{res}} \approx 0.1\left(\frac{m_2}{M_{\mathrm{Io}}}\right)^{1/3}\left(\frac{m_p}{8M_{\oplus}}\right)^{-1/3}.
    \label{eressmall}
\ee
Then from the eccentricity result, the average tidal heating luminosity temperature is estimated to be
\be
\begin{aligned}
    L_{\mathrm{res}} \approx \SI{2e16}{}\left(\frac{Q/k_2}{10^2}\right)^{-1}\left(\frac{R_1}{R_{\mathrm{Io}}}\right)^5\left(\frac{P_1}{2\, \mathrm{d}}\right)^{-5}\\ \times \left(\frac{m_2}{M_{\mathrm{Io}}}\right)^{2/3}  \left(\frac{m_p}{8\,M_{\oplus}}\right)^{-2/3} \mathrm{W}.
    \label{Lres}
\end{aligned}    
\ee
Finally, solving for the average temperature due to tides gives
\be
\begin{aligned}
    T_{\mathrm{res}} \approx 303\left(\frac{Q/k_2}{10^2}\right)^{-1/4}\left(\frac{R_1}{R_{\mathrm{Io}}}\right)^{3/4}\left(\frac{P_1}{2\, \mathrm{d}}\right)^{-5/4}\\ \times \left(\frac{m_2}{M_{\mathrm{Io}}}\right)^{1/6}  \left(\frac{m_p}{8\,M_{\oplus}}\right)^{-1/6} \mathrm{K}.
    \label{Tres}
\end{aligned}
\ee
Using this method, we estimate the inner moon temperature for the system in Figure~\ref{zeroecctides} to be 369.5\;K, within 2\% of the simulated result. The difference between the two values can again be attributed to the center of mass location that is offset from the origin. Indeed, this first estimate was for a planet with mass $m_p = 4.5\,M_\oplus$ ($m_p/m_2 \sim 300)$, and increasing the planet-moon mass ratio increases the agreement between the predicted and simulated tidal heating temperatures. Although the constant Hamiltonian configuration described by Equation~(\ref{Hamiltonian}) does not include tidal heating, the formulation described above provides an accurate picture of the short term heating rates. In Section~\ref{subsec:lifetime}, we consider evolution of the system over the longer term through damping timescales and lifetimes of heated moons.

We have also assumed that the inner moon is the test particle in this system as described in Section~\ref{3body} and thus use a massless particle to simulate it in Rebound. We have analyzed the effect this approximation would have on the dynamics by running similar simulations with an Io-mass moon instead. We find that there is less eccentricity perturbation and thus less tidal heating when using a massive inner moon, an effect that is greater for lower mass host planets. However, even in the case of a low mass rocky planet, the difference in tidal heating temperature is less than 20\%, and for the Jupiter-Io-Europa system, the difference we find is negligible.

In addition, the CPL tidal heating model is a simplified parameterized method of generalizing tidal interactions based on fixed values $Q$ and $k_2$. In detail, there is a complicated mechanism of heat transfer within a solid body that depends on parameters like the viscosity, shear modulus, and density of the material of the mantle, among others. The viscoelastic model of tidal heating aims to incorporate these variables \citep{dobosheller}. However, this is beyond the scope of this study and will be considered in future work (although for an example of the impact of using a viscoelastic moon model, see \citealp{rovira}).

Returning to the case of other MMR configurations, we tested other ratios such as 3:1 and 3:2 by running additional simulations and find that a moon in these MMR will incur a similar amount of heating, depending on the exact orbital periods of the moons. For example, moving the outer moon to a 3 day orbit in the system shown by Figure~\ref{zeroecctides} creates a 3:2 resonance and the inner moon is heated to 338\;K. In short, other resonant configurations can be similarly used to trace orbital evolution and tidal heating in exomoons.

\subsection{Tidal Heating of Io}

\begin{figure}
\epsscale{1.1}
\plotone{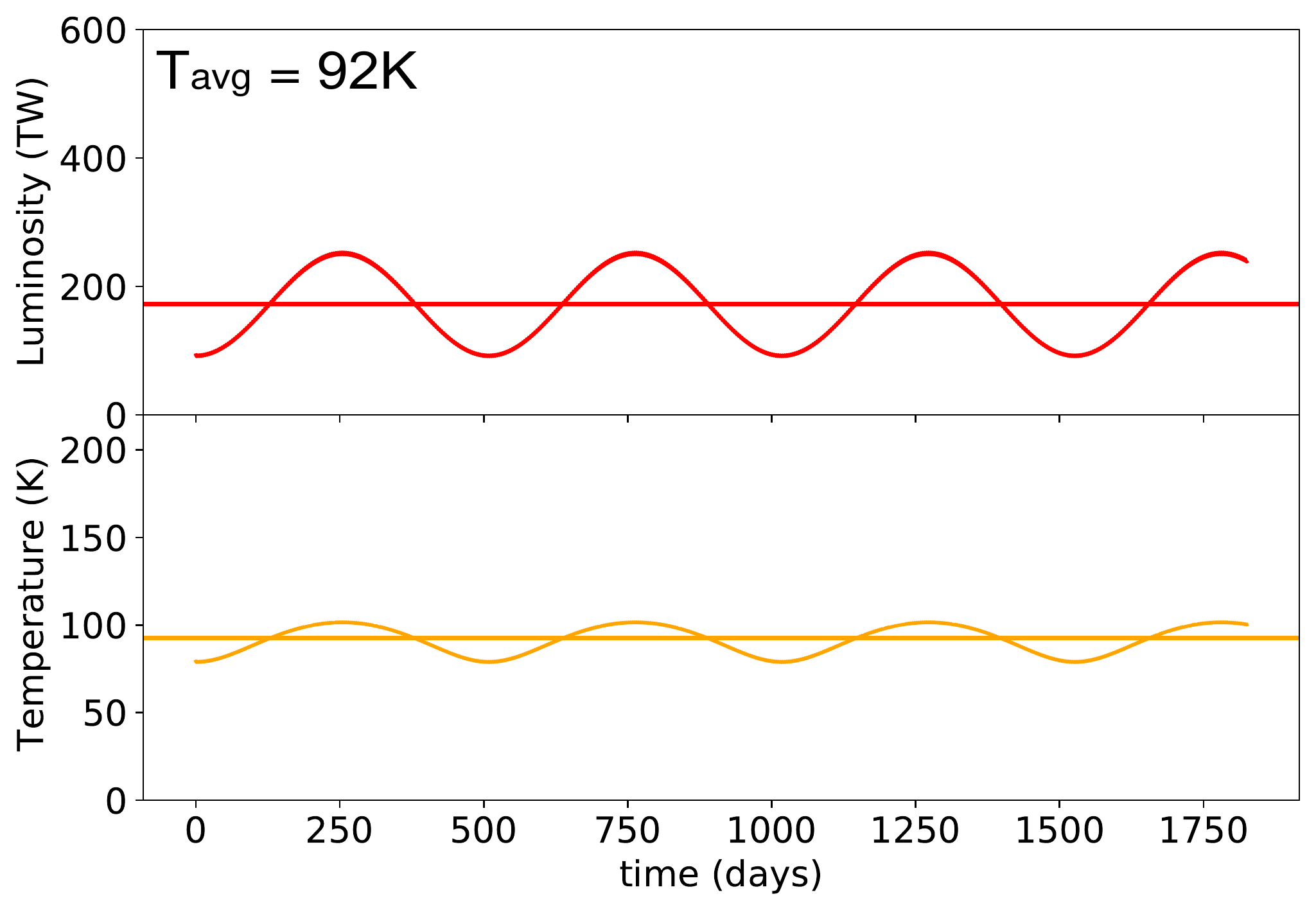}
\caption{Similar to Figure~\ref{zeroecctides} but for Io in the three body system consisting of Jupiter, Io, and Europa. The average tidal heating temperature of Io is 92K, which agrees with the upper bound set by observation.}
\label{heatingio}
\epsscale{1.0}
\end{figure}

Jupiter's moon Io is in a 2:1 resonance with the moon Europa making it a helpful test case for our heating models. Although the system is close to perfect resonance, the average eccentricity is significant enough to drive heating within Io. The observed tidal heat flux through Io is 2-4$\, \rm{Wm^{-2}}$ \citep{laineyk2}, corresponding to a surface temperature of $77-92$\,K \citep{tyler}. To check our heating model against observation, we simulate the three body system of Jupiter, Io, and Europa and obtain the oscillating tidal luminosity and temperature, which is averaged to obtain a $T_{\mathrm{avg}}$ of 92\,K (Figure~\ref{heatingio}). This agrees with the upper end of observation. Here we have used $k_2=0.54$ \citep{laineyk2} and $Q=36$ \citep{yoderQio} for Io. Note the difference in amplitude between the Io case and the zero eccentricity case; the tidal luminosity of Io varies relatively little, and this is because the resonance is close to perfect (the system is near a fixed point).

\begin{figure*}[htbp]
\epsscale{1.1}
\includegraphics[width=\textwidth,height=16cm]{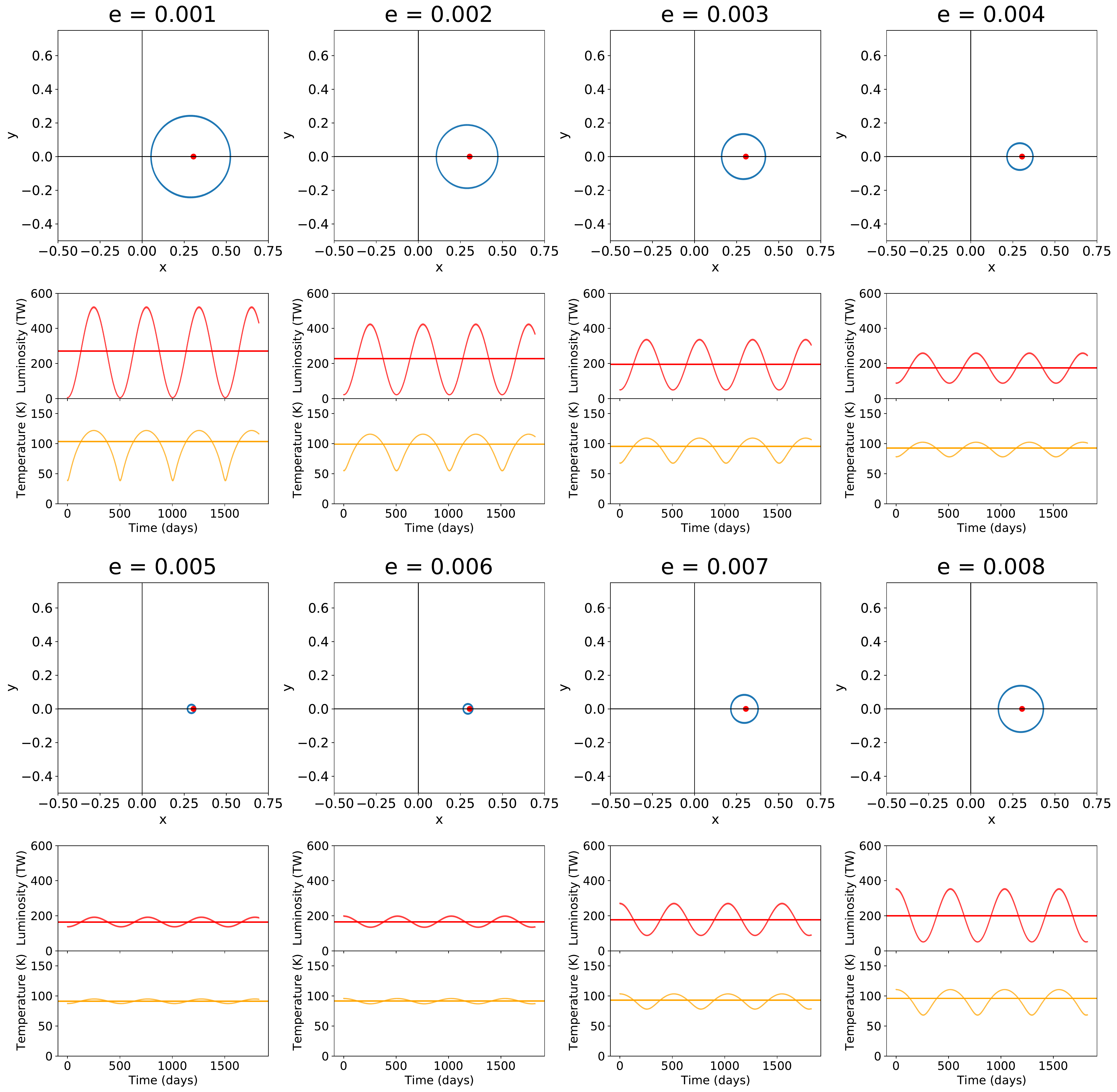}
\caption{Constant Hamiltonian trajectories and tidal heating curves for the Jupiter-Io-Europa system at varying eccentricities for Io's orbit. The blue curve in each shows the changing semimajor axis and eccentricity of Io and the red dot is the fixed point where this variation is minimized. The red curves are the tidal luminosity generated in Io and the orange curves below are the corresponding tidal temperature. The system is closest to perfect resonance at $e=0.005$, where the blue circle is smallest and the luminosity varies the least.}
\label{JupIoRes}
\epsscale{1.1}
\end{figure*}

We next explore the range of heating possible in a Jupiter-Io-Europa system to better understand how sensitive the heating is to the Hamiltonian value. We do this by varying initial eccentricity, starting from $e=0.001$ and incrementing by $0.001$ up to $e=0.008$ (Figure~\ref{JupIoRes}). The constant Hamiltonian curves circulate the fixed point for all cases. The radius of this circle is largest for $e=0.001$ and shrinks as we increase the eccentricity, getting close to the minimum at $e=0.005$. As we further increase $e$, the circle begins to expand again.

Figure~\ref{JupIoRes} also includes the heating luminosity and temperature for each value of eccentricity. The trend follows the pattern given by the trajectories of constant Hamiltonian: the variation in heating is a maximum for $e=0.001$, decreases as we raise $e$ until $e=0.005$, and then increases again. We can tell that the fixed point is between $e=0.001$ and $e=0.008$ also from the shape of the luminosity curve. At $t=0$, the luminosity is at a minimum for $e\leq0.005$ but at a maximum for $e\geq0.006$, signifying that there is minimum variation in heating somewhere between these eccentricities. Although the tidal heating temperature is lowest when we are closest to the fixed point (91\,K vs 108\,K for $e=0.001$) the difference between the maximum and minimum temperatures is less than 20$\%$, displaying that for a range of Hamiltonian values, and even perfect resonance, orbits can give rise to significant tidal heating.

\subsection{Composition and Period Dependency}

\begin{figure*}
\epsscale{1.1}
\includegraphics[width=\textwidth,height=16cm]{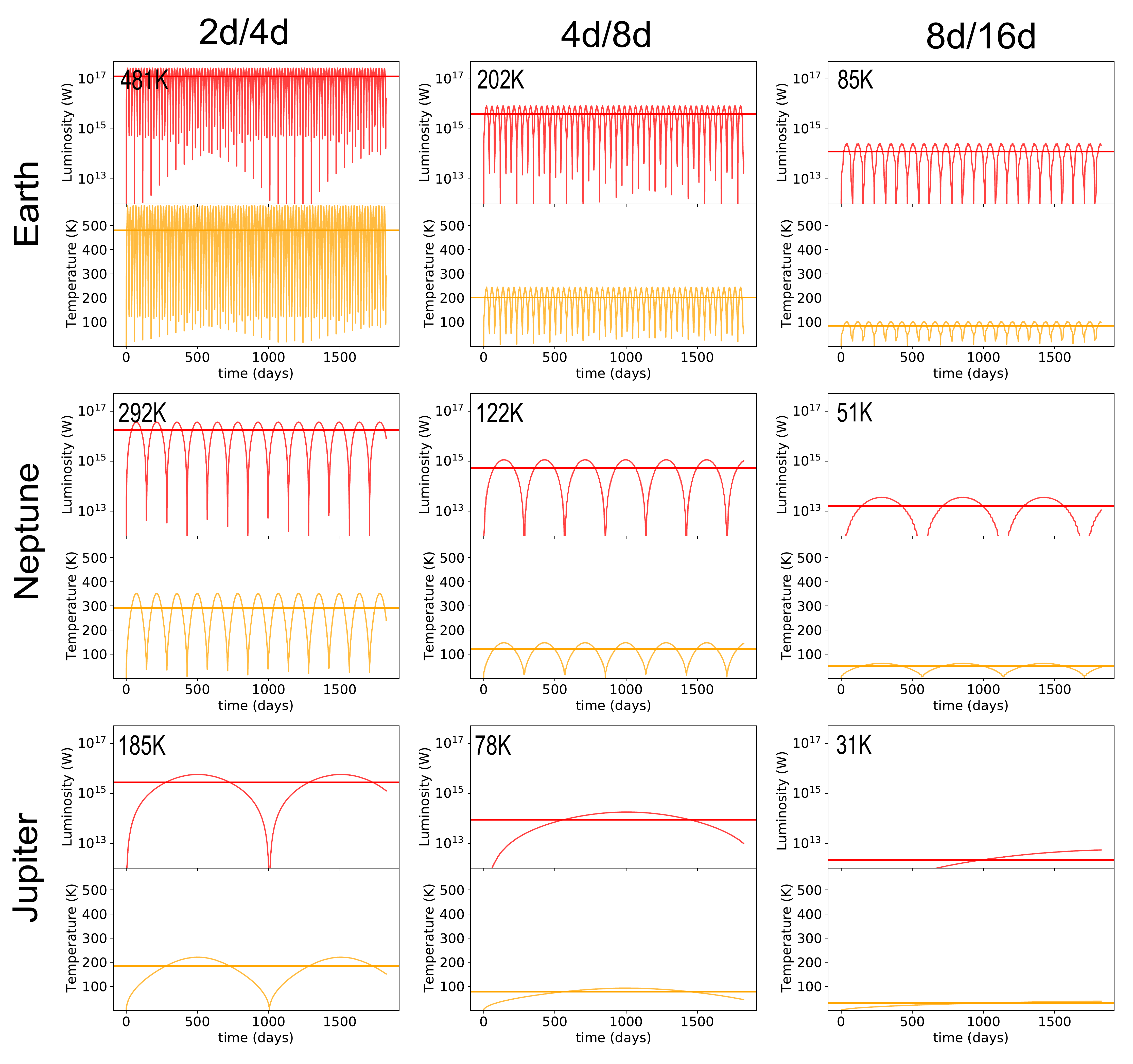}
\caption{Tidal heating luminosities and temperatures for two Io-like moons orbiting a planet in a 2:1 resonance. Each row corresponds to a different planet mass and each column refers to the orbital periods of the two moons in resonance. The red curves represent the tidal luminosity with the red horizontal line set at the average luminosity. The tidal heating temperature is shown by the orange curves, with the average temperature, the orange horizontal line, calculated using the average luminosity and displayed at the left corner of each plot. There is a clear trend of decreasing temperature as planet mass and orbital periods increase.}
\label{cascade}
\epsscale{1.1}
\end{figure*}

The results above beg the question: What kind of configuration leads to the most or the least heating within a moon? We next explore how central planet mass, and hence composition, and moon orbital period impact tidal heating within an orbiting moon. To study this, we simulate nine different systems with varying $m_p$ and $P_1$. We choose three different planet masses: $1\,M_{\oplus}$, $17\,M_{\oplus}$, and $318\,M_{\oplus}$, corresponding to Earth, Neptune, and Jupiter. We then choose three pairs of orbital periods for the moons: 2d/4d, 4d/8d, and 8d/16d. The inner moon is massless but has the same radius, Love number, and quality factor as Io, and the perturbing moon is identical except it has the mass of Io. Both moons are in initially circular orbits. Using the same method of calculating tidal luminosity over time, averaging, then finding $T_{\mathrm{avg}}$, we obtain the results for the nine combinations of these parameters (Figure~\ref{cascade}). Here, $K$ and $\Delta$ do not take on specific values but are determined by physical parameters as the system evolves. As implied by Equation~(\ref{Lres}), we get the most heating from low-mass planets and small orbital periods.

The maximum amount of heating is for an Earth-like planet with moons in orbit of two and four days. The temperature is 481\,K, even greater than the equilibrium temperature of Mercury! This highlights how strongly a moon can be tidally heated when it moves into a resonance -- even if it starts in a circular orbit. As the periods are increased across each row, the amount of heating dramatically decreases as expected from the strong scaling with period in Equation~(\ref{Lres}). Similarly, as planet mass is increased moving down each column, the moon-planet distance again increases to retain the same orbital periods and this drives heating down as well. Another effect evident from Figure~\ref{cascade} is the trend of libration time. As the orbital periods and planet mass increase, the oscillation period for luminosity and temperature increase as predicted by Equation~(\ref{tlib}).

\subsection{Lifetime of Tidally Heated Moons}
\label{subsec:lifetime}

The large amount of tidal heating found in the previous sections will sap energy from the moon's orbit and eventually move the moon out of resonance. The evolution of the mean motion, $n$, of the moon depends on the quality factor and Love number of both the planet, $Q_p$ and $k_{2,p}$, and the moon. Assuming the planet rotates at a faster rate than the moon orbits, and taking eccentricity to second order, this change in mean motion is given by \citet{boue}, 
\be
\begin{aligned}
   \frac{1}{n}\frac{dn}{dt}=-\frac{m_1}{m_p}\left(\frac{R_p}{a_1}\right)^5\frac{k_{2,p}}{Q_p}\left(\frac{9}{2}+\frac{459}{8}e^2\right)n
   \\
   +\, \frac{171}{2}\frac{m_p}{m_1}\left(\frac{R_1}{a_1}\right)^5\frac{k_{2}}{Q}e^2n.
    \label{dndt}
\end{aligned}
\ee
Similar to \citet{rovira}, we define a migration timescale of the moon 
\be
\frac{1}{t_{mig}}=-\frac{9}{2}\frac{m_1}{m_p}\left(\frac{R_p}{a_1}\right)^5\frac{k_{2,p}}{Q_p}n,
\label{tmig}
\ee
and an eccentricity damping timescale
\be
\frac{1}{t_{damp}}=\frac{21}{2}\frac{m_p}{m_1}\left(\frac{R_1}{a_1}\right)^5\frac{k_{2}}{Q}n,
\label{torb}
\ee
so that Equation~(\ref{dndt}) becomes
\be
\frac{1}{n}\frac{dn}{dt}=\frac{1}{t_{mig}}\left(1+\frac{51}{4}e^2\right)+\frac{57}{7}\frac{1}{t_{damp}}e^2.
\label{dndt2}
\ee

Thus when a two moon system migrates into a 2:1 resonance, we expect the phase of enhanced heating we find to last roughly this damping timescale. Applying Equation~(\ref{torb}) to the nine cases presented in Figure~\ref{cascade} provides a sense of the range of damping timescales. At the extremely short-lived end, a moon in a 2 day orbit around a $1M_{\oplus}$ planet peaks at 481\,K and will have its eccentricity damped in less than $1$\,Myr. In the other limit, the eccentricity of a moon in an 8 day orbit around a $1M_J$ planet that is heated to only 31\,K will last over $100$\,Myr. For many  of our most strongly heated cases, it may be difficult to find the moons in such a state unless a large number of planets are surveyed. That said, finding moons that are strongly heated will put strong constraints on the age and dynamical history of these systems.

Even if the moons move out of resonance, there is still the opportunity for the moons to migrate from tidal interaction with the planet, move back toward resonance, and reinvigorate the heating. To assess the steady-state heating for a moon that experiences this over secular timescales, we follow the arguments of \citet{rovira}. We summarize the rough order of magnitude conclusions here, but suggest the reader consult this work for a more detailed discussion.

First, for a two moon system, the inner moon must migrate outward faster than the outer moon to maintain resonance. Comparing these timescales for the two moons, and setting $2P_1=P_2$, we find that $m_2\lesssim20m_1$ for $t_{\mathrm{mig,}1}<t_{\mathrm{mig,}2}$. Second, we estimate the evolution over secular timescales by balancing eccentricity damping (which is trying to move the moons out of resonance) with the outward migration of the inner moon (which is trying to move the moons into resonance). Setting Equation~(\ref{dndt2}) to zero and solving for $e$ gives an equilibrium eccentricity
\be
\begin{aligned}
   e_{
   \mathrm{eq}} =0.055\left(\frac{Q}{Q_p}\right)^{1/2} \left(\frac{k_2}{k_{2,p}}\right)^{-1/2}\left(\frac{R_1}{R_{\mathrm{Io}}}\right)^{-5/2}\\ \times\left(\frac{R_p}{2R_{\oplus}}\right)^{5/2}  \left(\frac{m_1}{M_{\mathrm{Io}}}\right) \left(\frac{m_p}{8\,M_{\oplus}}\right)^{-1}.
    \label{eeq}
\end{aligned}    
\ee
Substituting this into Equation~(\ref{tidalL}) results in an equilibrium luminosity
\be
\begin{aligned}
    L_{\mathrm{eq}} =10^{16}\,\; \left(\frac{Q_p/k_{2,p}}{100}\right)^{-1}\left(\frac{R_p}{2R_{\oplus}}\right)^{5}\\ \times \left(\frac{P_1}{2\,d}\right)^{-5}   \left(\frac{m_1}{M_{\mathrm{Io}}}\right)^2  \left(\frac{m_p}{8\,M_{\oplus}}\right)^{-2} \, \mathrm{W}.
    \label{Leq}
\end{aligned}    
\ee
Such a luminosity could be expected for multi-moon systems that are continuously being driven into resonance by migration, representing a longer term steady-state tidal luminosity.

\section{Application to Exoplanets}
\label{sec:exoplanets}

Although there are many variables that impact the heating rate in exomoons around exoplanets, the above assessment showed that planet mass and moon orbital period are among the most significant. We next consider the population of known exoplanets \citep{NEA12}\footnote{Accessed on 2021-12-09} and discuss which are most conducive to hosting a detectable, tidally heated exomoon. This is not meant to be an exhaustive summary, but rather highlight the main considerations that should be used when trying to discover exomoons. We will also use some of the exoplanets we find here as concrete examples when we discuss detection methods in the following subsections.

\subsection{Key Exoplanet Parameters}
\label{sec:params}

Low mass rocky planets generate the most heat within an orbiting moon for a given period ratio (Equations (\ref{Lres}) and (\ref{Tres})). However, rocky planets are also the most dissipative, and their moons will undergo the quickest and most significant orbital evolution (\citealp{sasaki12,Tokadjian}). Furthermore, at least if we use our own solar system as an example, rocky planets may be less likely to harbor compact, multi-moon systems. Gas giants like Jupiter are the least dissipative so that their moons undergo less orbital evolution, but they also cause the least amount of heating. The ideal compromise between moon stability and temperature is thus found in moderate-sized planets like Neptune that have relatively low dissipation and give rise to appreciable tidal heating. These types of planets can range in size from just larger than super-Earths to ice giants. The former are called sub-Neptunes and have radii just over $1.7\,R_{\oplus}$ and masses typically less than $12\,M_{\oplus}$ \citep{lopez}. These types of planets give the best combination of low mass and slow damping timescale and will be favored moving forward.

\begin{figure}
\epsscale{1.1}
\plotone{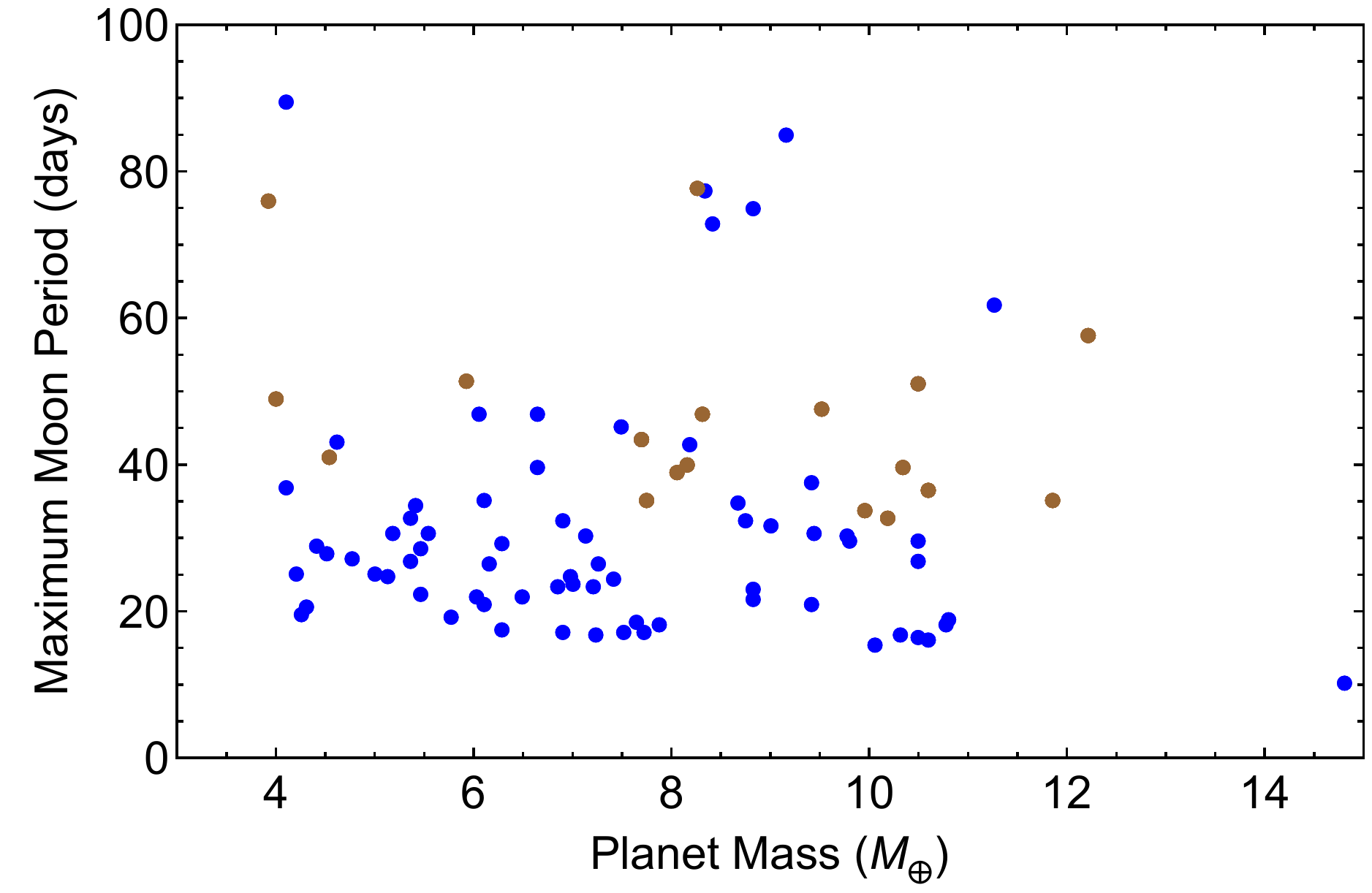}
\caption{Maximum orbital periods of exomoons around 25 rocky planets that can potentially retain a moon for over 300\,Myr (brown) and 85 Neptune-like planets that can retain a moon for over 5\,Gyr (blue). The smallest maximum moon period is 10 days and the largest is 90 days. This stability range allows for placement of moons in a variety of 2:1 orbital configurations, like the 2d/4d periods we consider in this paper.}
\label{safemoons}
\epsscale{1.0}
\end{figure}

Another important consideration is the range of periods where a moon can exist around a given planet. A moon that gets too close to the planet is subject to tidal dissipation and a moon that wanders too far away will be stripped away by the star (e.g.,\;\citealp{domingos}) and in many cases may eventually collide with planet \citep{hansen}. Starting from the list of all confirmed exoplanets, we categorize those with radius less than $2.4\,R_{\oplus}$ or mass less than $14\,M_{\oplus}$ as rocky Earth-like planets and those with radius $1.75\,R_{\oplus}$--$3.5\,R_{\oplus}$ or mass $4\,M_{\oplus}$--$14\,M_{\oplus}$ as Neptune-like, considering those planets that fall under both categories as each case separately. From this list, there are 25 rocky planets with a minimum moon retention time of 300\,Myr and 85 Neptune-like planets that can hold a moon for over 5\,Gyr. We use the method described in \citet{Tokadjian} to calculate the maximum moon periods for these 110 exoplanets (Figure~\ref{safemoons}), and set $Q_p=12$ for Earth-like planets and $Q_p=10^4$ for Neptune-like planets (\citealp{goldreich,murray}). The minimum moon period is set by the radius at which the moon is tidally disrupted by the planet, which depends on the moon's density \citep{Frank2002}. Since we choose an Io-sized moon for all calculations, the value for all cases is 0.24 days. The maximum moon period varies from about 10 days to 90 days, demonstrating that there is a wide range of orbital periods where a moon may exist for all of these planets. Motivated by the Jupiter-Io-Europa system, we choose orbital periods of 2 days and 4 days for moons in resonance for the remainder of this work. This ensures that the inner moon is close enough to the planet to incur significant tidal heating but far enough to maintain stability in the wake of orbital evolution. Using the scalings we derive analytically and numerically in Section~\ref{sec:heating}, these results can be extrapolated to other moon periods.

\begin{figure}
\epsscale{1.15}
\plotone{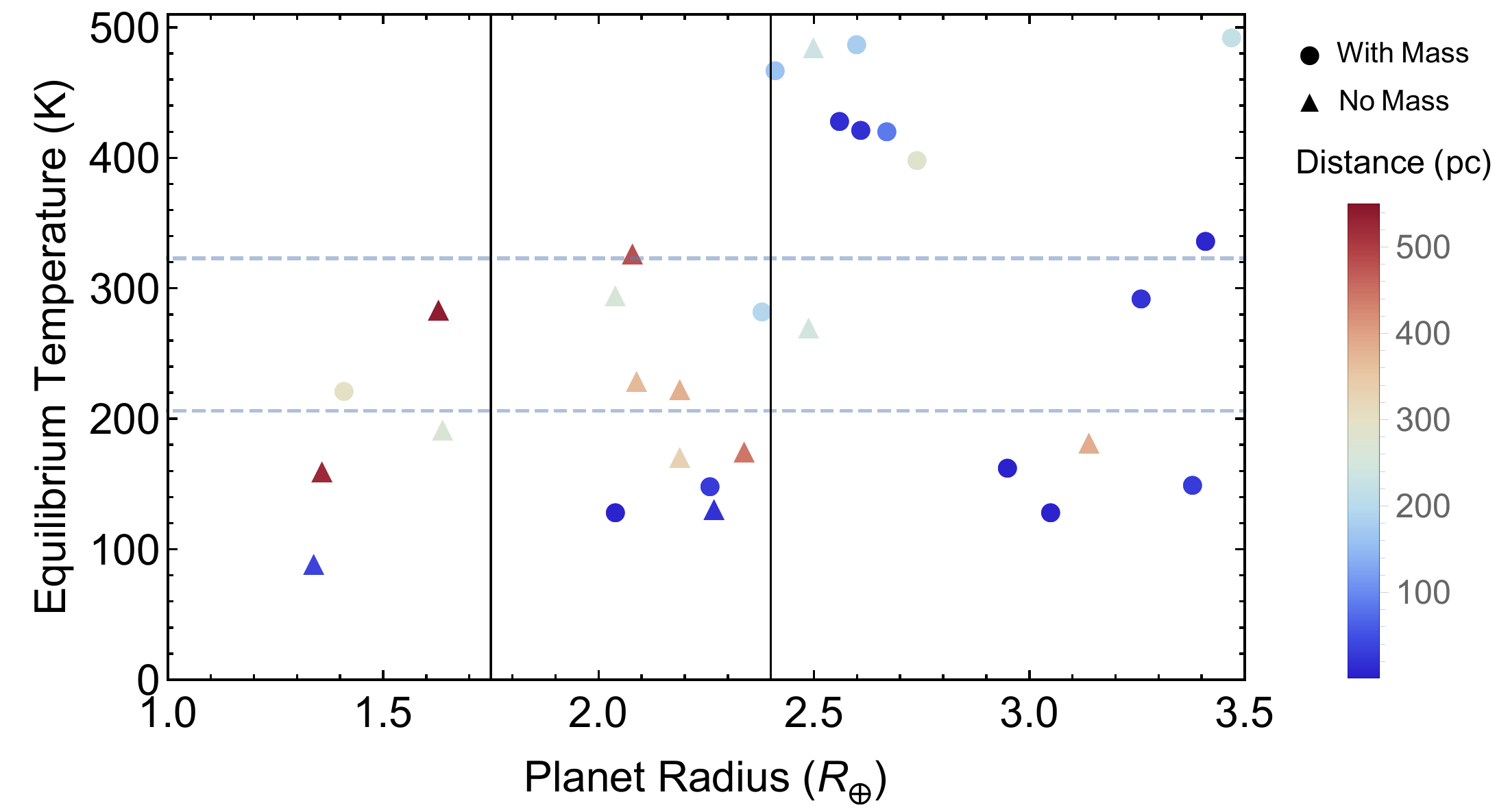}
\caption{Analysis of three properties of 30 exoplanets to distinguish potential exomoon detections through tidal heating. Each point traces $T_{\mathrm{irr}}$ of the given planet plotted against the planet's radius. The color corresponds to the distance from the Earth and filled circles are the planets with available estimated mass. The solid vertical lines divide the planets by composition with rocky on the left and Neptune-like on the right while the horizontal dashed lines represent the habitable zone. }
\label{planetcands}
\epsscale{1.0}
\end{figure}

When trying to identify planets where exomoon detection is possible, it is also important to consider the contrast between the temperatures of the moon and planet. As we will describe in the next subsection, a larger difference in temperature is helpful in observations. Thus for each exoplanet candidate, we estimate an irradiation temperature

\be
    T_{\mathrm{irr}} = \left(\frac{L_s}{16\pi a_p^2\sigma_{\mathrm{SB}}}\right)^{1/4},
    \label{Teq}
\ee
where $L_s$ is the luminosity of the parent star and $a_p$ is the semimajor axis of the planet. This temperature, along with the planet radius and distance to Earth, are compared for 30 exoplanets in Figure~\ref{planetcands}. These 30 planets are a sample from the list of planets in Figure~\ref{safemoons} made up of the 15 closest Neptune-like planets and 15 closest rocky exoplanets. The vertical lines split the plot into three segments which corresponds to planet composition: strictly rocky, rocky or Neptune-like, and strictly Neptune-like based on the radius cut described in \citet{fulton2017}. The horizontal lines represent the habitable zone, while circles and triangles show whether the planet has a determined mass or not, respectively. The color of each marker indicates the distance to the Earth. 

As described above, planets in the sweet spot for moon stability and tidal heating have the radius of a sub-Neptune (towards the right of the plot) and are nearby (blue markers). To maximize planet-moon temperature contrast, planets that have low mass and radius (towards the left of the plot), relatively cool in temperature (towards the bottom of the plot) and are nearby (blue markers) would be ideal. For planets with no estimated mass, a mass is assigned based on the radius and density cuts described in \citet{Tokadjian}.

Given the relative difficulty with detecting even rocky planets, finding the first rocky exomoon is a challenging task. Thus, novel and indirect methods may be necessary when searching for exomoons. This includes the secondary eclipse method and the search for sodium signatures which would result from volcanic outgassing. In the following, we examine the potential of these methods and follow up with a discussion of promising missions and instruments that may be applicable.

\subsection{Secondary Eclipse Method}
\label{sec:sececlipse}

We first study the method of detecting tidally heated exomoons via the technique of secondary transits. This depends on analyzing the relatively small dip in light as the star eclipses the planet. If the planet hosts a moon that is significantly heated, the inferred dayside planet luminosity would be larger than expected.

\begin{figure}
\epsscale{1.17}
\plotone{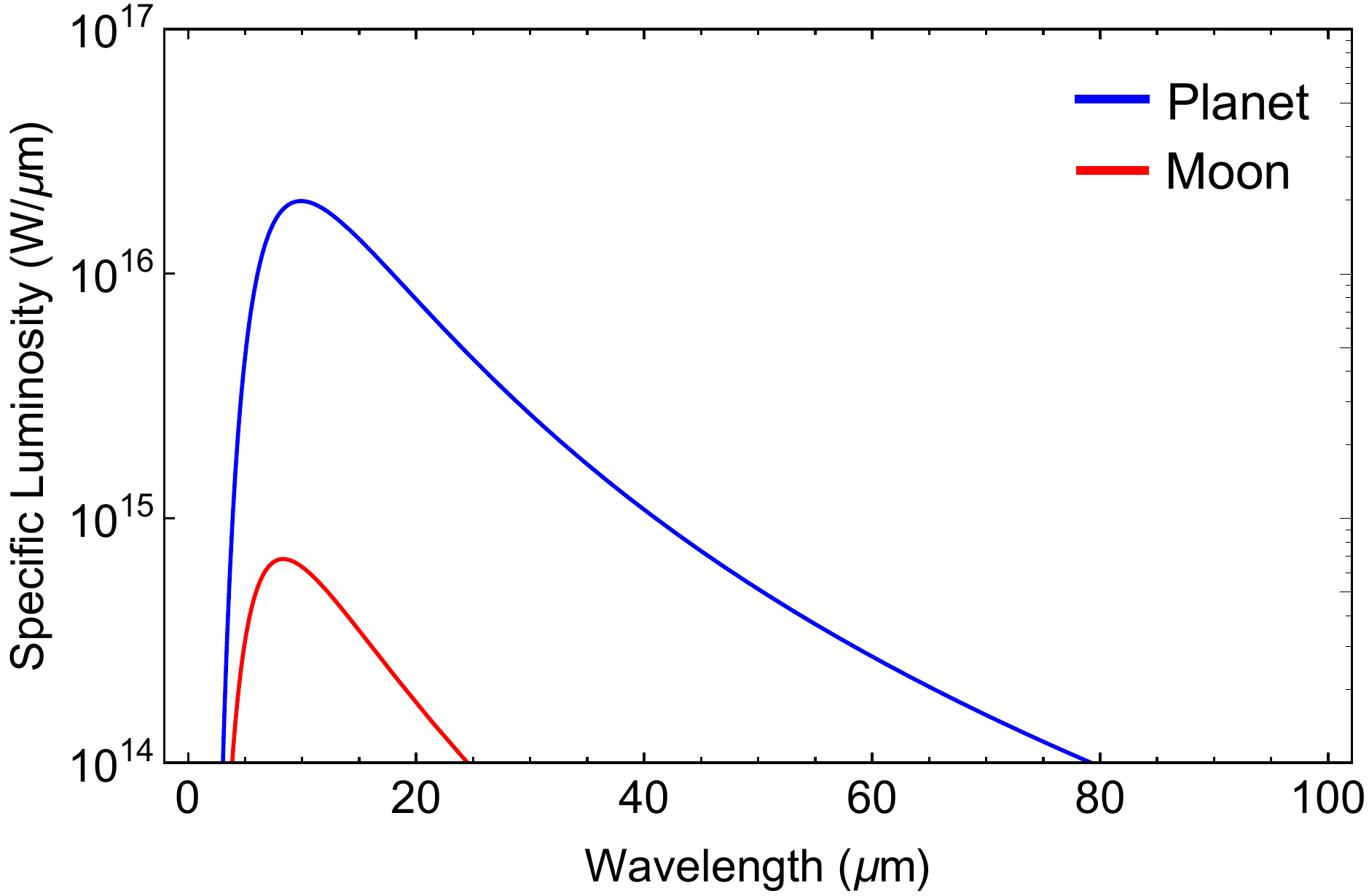}
\caption{Blackbody curve for K2-18b and a theoretical moon orbiting in a 2:1 resonance with an outer moon. $T_{\mathrm{irr}}=293$\,K and the tidal heating temperature of the moon is also 293\,K, giving $T_{\mathrm{tot}}=348$\,K for the moon, not nearly enough to match the luminosity of the planet.}
\label{K2-18bBB}
\epsscale{1.0}
\end{figure}

The amplitude of the secondary eclipse signal is given by the fractional change of light when a planet and moon, with luminosities of $L_p$ and $L_m$ respectively, go behind the star with luminosity $L_s$,

\be
r_t=\frac{L_p+L_m}{L_s} =  \frac{R_p^2T_{\mathrm{irr}}^4+R_1^2(T_{\mathrm{tide}}^4+T_{\mathrm{irr}}^4)}{R_s^2T_s^4},
\label{eqnratio}
\ee
where we take the equilibrium temperature for the planet using Equation~(\ref{Teq}). Thus, to maximize the signal of the moon in comparison to the planet, it is preferable to consider maximizing the ratios $R_1/R_p$ and $T_{\mathrm{tide}}$/$T_{\mathrm{irr}}$. As an example, the blackbody curves in Figure~\ref{K2-18bBB} show the specific luminosity of planet and tidally heated inner moon for K2-18b, a 8.88\,$M_{\oplus}$ planet at the cusp between super-Earths and sub-Neptunes. The system has an equilibrium temperature and moon heating temperature of 293\,K. It is apparent from this plot that the moon's luminosity is not enough to be detected apart from the planet because here $R_1 \ll R_p$ and $T_{\mathrm{tide}}\approx T_{\mathrm{irr}}$. Most of the sub-Neptunes in our sample face the same issue: because these planets have larger radii than rocky planets and incur relatively little tidal heating in an orbiting moon, the temperature contrast between planet and moon is not significant enough to separate the luminosity from the two bodies.

\begin{figure}
\epsscale{1.17}
\plotone{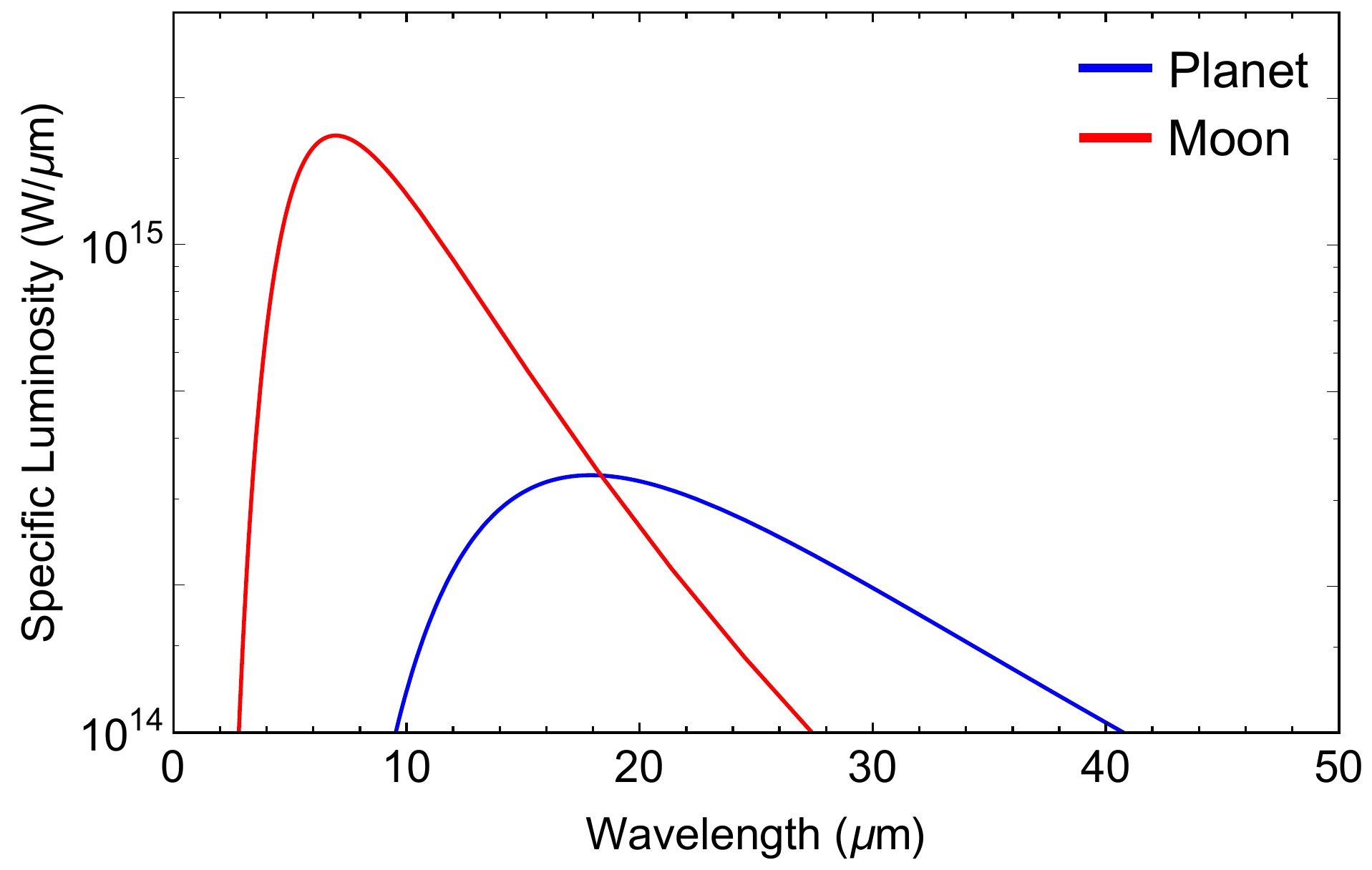}
\caption{Blackbody curve for TOI-4353.01 and a theoretical Io-sized moon orbiting in a 2:1 resonance with an outer moon. In this case, the moon outshines the planet because of significant tidal heating which makes this planet a good candidate for possible moon detection through the secondary eclipse method. At the wavelength where the moon's curve peaks, the moon's luminosity is more than 100 times the planet's luminosity.}
\label{GJ3323cBBb}
\epsscale{1.0}
\end{figure}

On the other hand, the signal of the moon can be much larger when the parent planet is a smaller sub-Neptune or a rocky Earth-like planet. In particular, we find the light curves of planet and moon begin to separate when the ratio between the moon's peak luminosity and the planet's luminosity at that wavelength is greater than a factor of two. Table~\ref{table1} summarizes this luminosity ratio along with $T_{\mathrm{res}}$ (Equation~(\ref{Tres})), distance from the Earth, and planet type for 15 exoplanets with the highest luminosity ratio of the 30 shown in Figure~\ref{planetcands}. Also listed is the moon-planet luminosity ratio using $L_{eq}$ (Equation~(\ref{Leq})), which is relevant for longer term tidal heating. For the secondary eclipse method, we focus on the short term heating rates where the moon's luminosity may be much greater than the planet's. The planet with the highest moon-to-planet luminosity ratio is TOI-4353.01, where the moon outshines the planet by two orders of magnitude at the peak wavelength of the heated moon. Figure~\ref{GJ3323cBBb} shows the blackbody plots for this case: the red curve traces the moon's luminosity which is clearly discrete from the blue curve of the planet. The tidal heating temperature of the moon is 414\,K so that the total temperature of the moon, irradiation plus tidal heating, is 417\,K. Using Equation~(\ref{eqnratio}) for TOI-4353.01 we get $r_t=0.026$\,ppm.

\begin{deluxetable}{lcccccc}
\tablecolumns{7} 
\tablewidth{125pt}
\vspace{0.5cm}
\tablecaption{15 Exoplanets With Largest $L_1/L_p$ \label{table1}}
\tablehead{
Planet& $T_{\mathrm{res}}$ & $\lambda_{1,\mathrm{peak}}$ & $L_{1,\mathrm{res}}/L_p$ &$L_{1,\mathrm{eq}}/L_p$ & Distance &Type\\
&($\mathrm{K}$)&($\mu m$)&&&(pc)
}
\startdata
TOI-4353.01 & 414 & 6.96 & 109 & 0.61 & 25.1 & Rocky\\
Wolf-1061 d & 338 & 8.54 & 54 & 0.63 & 4.31 & Rocky\\
Wolf-1061 d & 338 & 8.54 & 24 & 0.17 & 4.31 & Neptune\\
TOI-4191.01 & 320 & 9.01 & 18 & 0.37 & 83.9 & Rocky\\
TOI-4328.01  & 354 & 8.10 & 7.9 & 0.18 & 25.0 & Rocky\\
Kepler-1630 b & 326 & 8.73 & 5.6 & 0.04 & 331 & Rocky\\
GJ 3138 d & 320 & 8.94 & 4.8 & 0.16 & 28.5 & Rocky\\
Kepler-441 b & 377 & 7.56 & 3.9 & 0.11 & 268 & Rocky\\
Kepler-62 f & 401 & 7.07 & 2.6 & 0.08 & 301 & Rocky\\
Kapteyn c & 343 & 8.34 & 2.3 & 0.04 & 3.93 & Neptune\\
GJ 3138 d & 320 & 8.94 & 2.0 & 0.16 & 28.5 & Neptune\\
TOI-4349.01 & 315 & 9.00 & 0.88 & 0.06 & 72.5 & Rocky\\
Kepler-1536 b & 333 & 8.53 & 0.56 & 0.02 & 387 & Neptune\\
TOI-4503.01 & 333 & 8.26 & 0.25 & 0.03 & 62.5 & Rocky\\
Kepler-452 b & 378 & 7.14 & 0.25 & 0.03 & 552 & Rocky
\enddata
\vspace{-0.5cm}
\end{deluxetable}

\begin{figure}
\epsscale{1.15}
\plotone{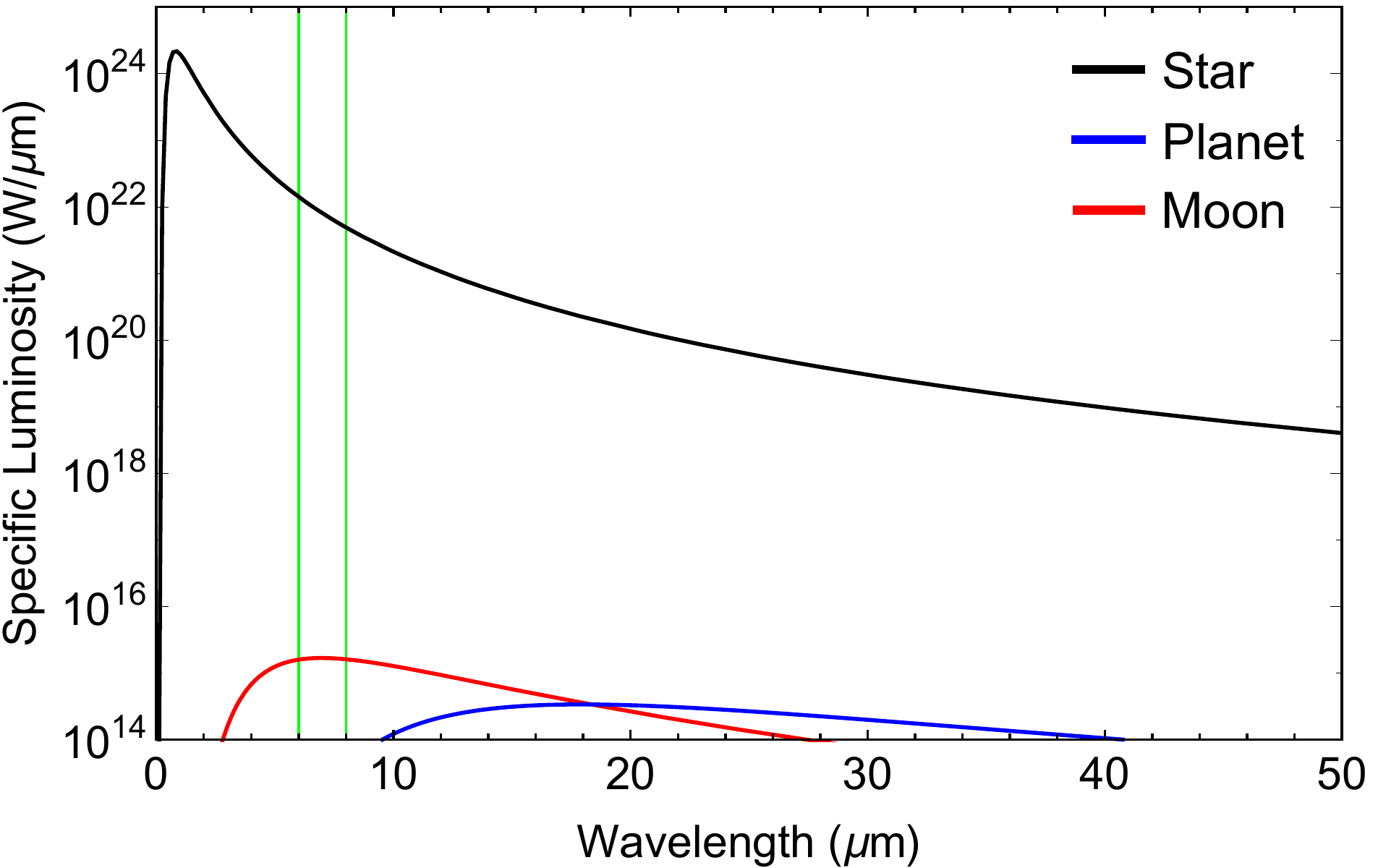}
\caption{Similar to Figure~\ref{GJ3323cBBb}, but including the luminosity of the star. The best wavelength range to observe the system in terms of detecting this moon is shown as green vertical lines, and corresponds to 6-8\,$\rm{\mu m}$.}
\label{GJ 3323 c}
\epsscale{1.0}
\end{figure}

However, this signal can be made even larger by focusing on a narrower band of wavelengths. By integrating over Planck's Law and normalizing by the disk area, similar to the method described in \citet{charb}, the precision for luminosity over wavelength range $\lambda_1$ to $\lambda_2$ becomes

\be
r_{\lambda} =  \frac{\int_{\lambda_1}^{\lambda_2} \left[ R_p^2B_{\lambda}(T_{\mathrm{irr}})+R_1^2\left(B_{\lambda}(T_{\mathrm{tide}})+B_{\lambda}(T_{\mathrm{irr}})\right)\right] \,d\lambda}{R_s^2\int_{\lambda_1}^{\lambda_2} B_{\lambda}(T_s) \,d\lambda}.
\label{eqnrangeprec}
\ee
Here $B_{\lambda}(T)$ is the spectral radiance given by Planck's Law integrated over half the hemisphere to give the specific luminosity. In the case of TOI-4353.01, the moon's radiation peaks at 6.96\,$\rm{\mu m}$, so observations in the range of 6-8\,$\rm{\mu m}$ would be ideal to collect the most light from the moon as possible (Figure~\ref{GJ 3323 c}). Applying Equation~(\ref{eqnrangeprec}), we obtain $r_{\lambda}=0.419$\,ppm.

\subsection{Sodium Signatures of a Heated Exomoon}

Volcanic vents on a tidally heated moon will eject certain kinds of elements such as sodium and potassium, whose clouds may be detectable via transit spectroscopy. In fact, Io is currently losing mass at a rate of about $\dot{M}_{\mathrm{Io}}=1000$\,kg s$^{-1}$, a fraction of which is neutral sodium that has been shown to be detectable through sodium line spectroscopy \citep{oza}. Similarly, observations of exoplanets with a tidally heated moon that is venting sodium at a similar or higher rate may provide evidence of the exomoon. Following the methods detailed in \citet{oza}, the average column density of sodium viewed against the background star is

\be
N_{\mathrm{Na}}=\frac{\dot{M_1}\tau}{m_{\mathrm{Na}}\pi R_s^2},
\label{coldensity}
\ee
where $m_{\mathrm{Na}}$ is the mass of a sodium atom, $R_s$ is the stellar radius, $\tau$ is the Na I lifetime, and $\dot{M_1}$ is the mass loss rate of the inner moon due to volcanism. We expect this mass loss rate to roughly scale proportional to the energy input from tidal heating. Thus, we scale to Io, giving

\be
    \dot{M_1} = \frac{L_1}{L_{\mathrm{Io}}}\dot{M_{\mathrm{Io}}}.
\ee
Here $L_1$ and $L_{\mathrm{Io}}$ are the tidal luminosities of the inner moon around the exoplanet and Io, respectively, where the latter is $1.25\,\rm{x}\,10^{14}\,\rm{W}$ \citep{veeder}. When we evaluate the damping rates for hypothetical moons, we find that they are very short, $\sim 1$\,Myr, for some of the most strongly heated cases. This means it may be difficult to build up a sufficiently large cloud of volcanic material and it is unlikely to be present long enough to have a chance of detection. For this reason, we focus on cases where resonance is maintained by migration and use $L_{\mathrm{eq}}$, given by Equation~(\ref{Leq}), for estimating $L$, where $Q_p/k_{2,p}\approx 300$ for Earth-like planets and $Q_p/k_{2,p}\approx 3000$ for Neptune-like planets. Table~\ref{table2} summarizes the 15 systems with the highest mass-loss rate from the list of nearest and most stable exoplanets. We indicate $\dot{M_1}$ in both $\mathrm{kg\;s}^{-1}$ and relative to Io and find that many of these tidally heated moons in resonance are likely to vent sodium at a rate that is tens of times that of Io.

\begin{deluxetable}{lcccc}
\tablecolumns{5} 
\tablewidth{120pt}
\vspace{0.5cm}
\tablecaption{15 Exoplanets With Largest Mass-Loss Rate \label{table2}}
\tablehead{
Planet& $\dot{M_1}$ $(\mathrm{kg\;s}^{-1})$ & $\dot{M_1}/\dot{M_{\mathrm{Io}}}$& Distance (pc) &Type
}
\startdata
TOI-4353.01 & \SI{1.18e5}{} & 118 & 25.1 & Rocky\\
Kepler-62 f & \SI{1.11e5}{} & 111 & 301 & Rocky\\
Kepler-452 b & \SI{0.98e5}{} & 98.4 & 552 & Rocky\\
Kepler-441 b & \SI{0.98e5}{} & 98.1 & 268 & Rocky\\
TOI-4328.01 & \SI{0.86e5}{} & 86.1 & 25 & Rocky\\
TOI-2091.01 & \SI{0.78e5}{} & 78.6 & 70.2 & Rocky\\
Kepler-1040 b & \SI{0.77e5}{} & 76.8 & 482 & Rocky\\
TOI-4503.01 & \SI{0.76e5}{} & 75.9 & 62.5 & Rocky\\
Kepler-1540 b & \SI{0.60e5}{} & 60.0 & 245 & Neptune\\
Kepler-1536 b & \SI{0.48e5}{} & 47.4 & 387 & Neptune\\
HD 216520 c & \SI{0.43e5}{} & 42.6 & 19.6 & Neptune\\
Kepler-126 d & \SI{0.36e5}{} & 35.7 & 237 & Neptune\\
Kapteyn c & \SI{0.33e5}{} & 33.6 & 3.93 & Neptune\\
Wolf-1061 d & \SI{0.26e5}{} & 25.5 & 4.31 & Neptune\\
HD 136352 d & \SI{0.22e5}{} & 21.9 & 14.7 & Neptune
\enddata
\vspace{-0.5cm}
\end{deluxetable}

\subsection{Detection with Future Satellites and Instruments }
Future missions may be able to detect exomoons through the indirect methods we present, but must have the necessary sensitivity and a well-matched wavelength range. For example, the planned satellite mission PLAnetary
Transits and Oscillations of stars (PLATO) \citep{plato} will have a photometric precision of $50\,\rm{ppm}\,\rm{hr}^{1/2}$, which is too large to detect this secondary eclipse of the system TOI-4353.01 \citep{marchiori}. In addition, PLATO will observe at a wavelength range of 0.5-1\,$\rm{\mu m}$, and $r_{\lambda}$ is much too small for this range to be detectable. However, this wavelength range includes the strong sodium doublet at 589\,nm so that sodium line spectroscopy with PLATO would be possible.

The James Webb Space Telescope (JWST) is able to perform spectroscopy using a mid-infrared instrument for wavelength range 0.6-29\,$\rm{\mu m}$, which is just outside the sodium doublet wavelength but fits the ideal case for the secondary eclipse of TOI-4353.01 \citep{jwst}. However, with a median precision of $170$\,ppm for wavelengths less than 10\,$\rm{\mu m}$, JWST is not as precise as PLATO \citep{venot} and would not be able to detect the secondary eclipse of TOI-4353.01. Varying potential exomoon size, resonance order, and tidal heating models should be explored in future work to better estimate the prospects for this telescope to find an exomoon.

Similarly, the Atmospheric Remote-sensing Infrared
Exoplanet Large-survey (ARIEL) will be sensitive to infrared radiation between 2\,$\rm{\mu m}$ and 7.8\,$\rm{\mu m}$, the edge of which corresponds to the peak of the test moon around TOI-4353.01 \citep{ariel}. With an estimated photometric precision of 10-100\,ppm, ARIEL will have sensitivity comparable to PLATO, and thus might be the best option to date in terms of detecting a tidally heated exomoon around an exoplanet using secondary transits.

Other space telescopes like the CHaracterising ExOPlanets
Satellite (CHEOPS) and the planned Roman Space Telescope (formerly WFIRST) have similar wavelength ranges as PLATO and likely will not find exomoons using the technique of secondary eclipses (\citealp{cheops}; \citealp{wfirst}). Nevertheless, these missions are set to answer critical questions about extraterrestrial habitability with techniques like microlensing to discover the smallest planets to date. Smaller planets would incur more heating in their moons for a given resonance mode, but these moons will likely be smaller in size and thus less luminous. Therefore, even with an ideal case of a significantly heated Io-sized moon, it would be difficult to reach the precision needed to detect such a moon with the secondary eclipse method.

Future ground based observatories such as the Giant Magellan Telescope (GMT), Extremely Large Telescope (ELT), and Thirty Meter Telescope (TMT) will all have instruments that can be used in a wide range of infrared astronomy (\citealp{gmt}; \citealp{elt}; \citealp{tmt}). Given that the atmosphere's infrared window is at slightly longer wavelengths ($\sim$8-14\,$\rm{\mu m}$ vs a peak of 7.35\,$\rm{\mu m}$), the exomoon in question would need to have a lower overall temperature. This is more likely for larger exoplanets, like the sub-Neptune class mentioned earlier, so long as the equilibrium temperature of the bodies is small enough for the moon to still outshine its parent planet. Because these observatories will also observe in the optical, sodium line spectroscopy of potential tidally heated moons is promising.

\section{Conclusion}
\label{sec:conclusion}

In this study, we explored the tidal heating of resonant exomoons and the implications for moon detection. We first introduced the dynamics of orbital resonance, where we focused on the equations of the Hamiltonian, minimizing function, and libration time. We then explored more details of the dynamics with simulations of this 2:1 resonance using REBOUND and showed the agreement between theory and numerics. This demonstrated that even a moon with zero initial eccentricity can have its eccentricity pumped to increasingly large values when perturbed by an outer moon.

Since eccentricity, along with semimajor axis, affect tidal heating, the libration of exomoons near mean motion resonance cause oscillations in tidal luminosities. We showed that when such theory is applied to the Jupiter-Io-Europa system, we are able to accurately replicate Io's tidal heating temperature. We then explored the broader parameter space of exoplanet composition and masses, and exomoon periods, to understand the range of tidal heating possible. This demonstrated that rocky planets orbited by moons in a tight orbit will lead to the most tidal heating.

In some of the strongest heated cases, the resonance may be short lived. This would make it difficult to catch the moons when they are most strongly heated unless large surveys are used to probe many planets for moons. Nevertheless, we show that in some cases outward tidal migration can work to keep the moon in resonance, prolonging the time over which there is significant heating.

We then applied the heating model we developed to real exoplanet systems. This exercise demonstrated what considerations should be taken into account when assessing which systems are best suited for finding tidally heated exomoons.

We studied the observational implications of heated exomoons using these two methods. We first explored the secondary eclipse technique, where a planet's day side would be inferred to be too hot due to the presence of a heated moon. We generally found that this would be difficult to perform in practice because most instruments have insufficient sensitivity or do not have long enough wavelength coverage. The second method involved detecting sodium clouds on the parent planet fueled by the outgassing of its moon as it is heated. We found that the mass-loss rate of many of these moons may be tens of times larger than the current mass-loss rate of Io and thus may be detectable with PLATO or ground-based observatories whose instruments cover the sodium doublet wavelength range. A closer analysis of these planets and others by expanding the parameter space will better show which telescopes are best suited for potentially detecting a tidally heated exomoon in resonance around an exoplanet.

\acknowledgments
We acknowledge that support for this work was provided by the 2021 Carnegie Institution for Science Venture Grant program. A.T. acknowledges support from the USC-Carnegie fellowship.

\bibliographystyle{yahapj}

\end{document}